\documentclass[aps,prl,twocolumn,amsmath,amssymb,nofootinbib,superscriptaddress]{revtex4-1}

\usepackage{etoolbox}

\usepackage{relsize}
\usepackage{wasysym}
\usepackage{times}
\usepackage[pdftex]{graphicx}
\usepackage{dcolumn}
\usepackage{bm}
\usepackage{amsmath}
\usepackage{indentfirst}
\usepackage{float}
\usepackage[colorlinks]{hyperref}
\usepackage[dvipsnames]{xcolor}
\usepackage{xcolor}
\newcommand{\hhl}[1]{{\color{black}{#1}}}

\usepackage{subfigure}
\usepackage{algorithm}
\usepackage{algorithmicx}
\usepackage{algpseudocode}
\usepackage{amsmath}
\usepackage{verbatim}
\usepackage{makecell}
\usepackage{braket}
\usepackage{multirow}
\usepackage{textcomp}
\usepackage{colortbl}
\newcommand{\HFRC}[1][Hefei National Research Center for Physical Sciences at the Microscale and School of Physical Sciences, University of Science and Technology of China, Hefei 230026, China]{\affiliation{#1}}
\newcommand{\SHRC}[1][Shanghai Research Center for Quantum Science and CAS Center for Excellence in Quantum Information and Quantum Physics, University of Science and Technology of China, Shanghai 201315, China]{\affiliation{#1}}
\newcommand{\HFNL}[1][Hefei National Laboratory, University of Science and Technology of China, Hefei 230088, China]{\affiliation{#1}}
\newcommand{\HNKL}[1][Henan Key Laboratory  of  Quantum Information and Cryptography, Zhengzhou, Henan 450000, China]{\affiliation{#1}}
\newcommand{\CASTP}[1][CAS Key Laboratory for Theoretical Physics, Institute of Theoretical Physics, Chinese Academy of Sciences, Beijing 100190, China]{\affiliation{#1}}
\newcommand{\JIQT}[1][Jinan Institute of Quantum Technology and Hefei National Laboratory Jinan Branch, Jinan 250101, China]{\affiliation{#1}}
\newcommand{\NIM}[1][National Institute of Metrology, Beijing 102200, China]{\affiliation{#1}}
\newcommand{\QCTek}[1][QuantumCTek Co., Ltd., Hefei 230026, China]{\affiliation{#1}}
\newcommand{\SMX}[1][School of Microelectronics, Xidian University, Xi’an, China]{\affiliation{#1}}

\begin{document}

\title{Establishing a New Benchmark in Quantum Computational Advantage with 105-qubit \textit{Zuchongzhi}~3.0 Processor}

\author{Dongxin Gao} 
\thanks{These authors contributed equally to this work.}
\HFRC
\SHRC
\HFNL
\author{Daojin Fan} 
\thanks{These authors contributed equally to this work.}
\HFRC
\SHRC
\HFNL
\author{Chen Zha}
\thanks{These authors contributed equally to this work.}
\HFRC
\SHRC
\HFNL
\author{Jiahao Bei}
\SHRC
\author{Guoqing Cai}
\SHRC
\author{Jianbin Cai}
\HFRC
\SHRC
\HFNL
\author{Sirui Cao}
\HFRC
\SHRC
\HFNL
\author{Xiangdong Zeng}
\HFRC
\author{Fusheng Chen}
\HFRC
\SHRC
\HFNL
\author{Jiang Chen}
\SHRC
% ...



\begin{abstract}\relscale{1}
In the relentless pursuit of quantum computational advantage
, we present a significant advancement with the development of \textit{Zuchongzhi} 3.0. This superconducting quantum computer prototype, comprising 105 qubits, achieves high operational fidelities, with single-qubit gates, two-qubit gates, and readout fidelity at 99.90\%, 99.62\% and 99.18\%, respectively. Our experiments with an 83-qubit, 32-cycle random circuit sampling on \textit{Zuchongzhi} 3.0 highlight its superior performance, achieving one million samples in just a few hundred seconds. This task is estimated to be infeasible on the most powerful classical supercomputers, Frontier, which would require approximately $6.4 \times 10^9$ years to replicate the task. This leap in processing power places the classical simulation cost six orders of magnitude beyond Google's SYC-67 and SYC-70 experiments [Nature \textbf{634}, 328 (2024)], firmly establishing a new benchmark in quantum computational advantage. Our work not only advances the frontiers of quantum computing but also lays the groundwork for a new era where quantum processors play an essential role in tackling sophisticated real-world challenges.

\end{abstract}

\maketitle

\section{Introduction}

The quest for quantum computational advantage, a concept first coined by John Preskill, has been a central driving force in the field of quantum computing~\cite{preskill2018quantum,Wu2021strong,zhu2022quantum,arute2019quantum,morvan2023phase,zhong2020quantum,deng2023gaussian,zhong2021phase,madsen2022quantum}. This term captures the pivotal moment when a quantum computer is capable of executing calculations that are beyond the reach of even the most advanced classical computers. In 2019, Google claimed that its quantum processor, Sycamore~\cite{arute2019quantum}, had achieved this milestone, demonstrating its capabilities through an experiment on a problem known as random circuit sampling, which was deliberately designed to be intractable by classical systems. Random circuit sampling~\cite{boixo2018characterizing,bouland2019complexity,aaronson2016complexity} involves the creation of quantum states by applying a series of random quantum gates, followed by measurement. This process has become a focal point of intensive research due to its capacity to underscore the computational superiority of quantum systems. Since that breakthrough, the field has been eager to extend these results to larger systems and more complex tasks, striving to solidify the quantum advantage in computational tasks. Particularly in experimental settings, China's \textit{Zuchongzhi} processor~\cite{Wu2021strong,zhu2022quantum,Gong2021quantum,cao2023generation,zhao2022realization,ye2023logical} has been in continuous competition with Sycamore, pushing the boundaries of what is achievable with quantum technologies. To date, the largest scale of random quantum circuit is achieved by Google with 67 qubits at 32 cycles (SYC-67) and 70 qubits at 24 cycles (SYC-70)~\cite{morvan2023phase}.

In this work, we aim to challenge this record. We have developed \textit{Zuchongzhi} 3.0, a more powerful superconducting quantum computer prototype, equipped with 105 qubits and exceptionally high-fidelity manipulation capabilities. The single-qubit gate, two-qubit gate, and readout fidelities are 99.90\%, 99.62\%, and 99.18\%, respectively. Leveraging this prototype, our experiments utilize a significantly larger quantum circuit of 83 qubits at 32 cycles, thereby pushing the limits of current quantum hardware capabilities. On our \textit{Zuchongzhi} 3.0, the task of obtaining one million samples is accomplished in just a few hundred seconds. It is a stark contrast to the estimated $6.4 \times 10^9$ years, required by the most formidable supercomputers of today, Frontier, to replicate this sampling endeavor. Compared to Google's latest experiment, SYC-67 and SYC-70~\cite{morvan2023phase}, the classical simulation cost of our \hhl{83-qubit, 32-cycle} experiment is six orders of magnitude higher. Through this achievement, we establish a new benchmark in quantum computational advantage, which is essential for harnessing the full potential of quantum computing. Beyond this, our work opens avenues for investigating how increases in qubit count and circuit complexity can enhance the efficiency in solving real-world problems.

\section{\textit{\textit{Zuchongzhi}} 3.0 Quantum Processor}

\begin{figure*}[!htbp]
\begin{center}
\includegraphics[width=1\linewidth]{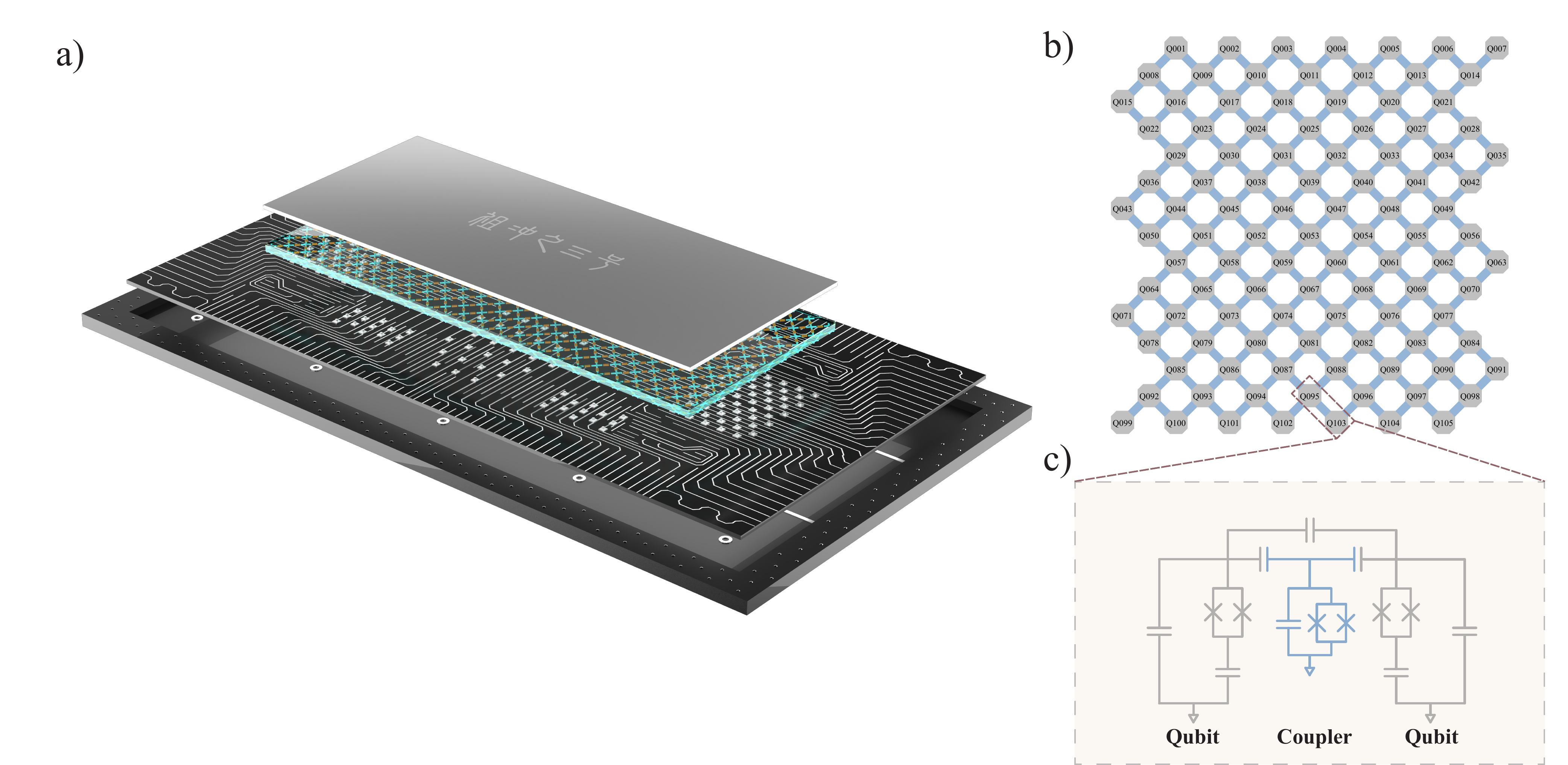}
\end{center}
\caption{\textbf{\textit{Zuchongzhi} 3.0  quantum processor. a)} The illustration of the \textit{Zuchongzhi} 3.0 quantum processor. The device consists of two sapphire chips integrated using a flip-chip technique. One chip integrates 105 qubits and 182 couplers, while the other is integrated with all the control lines and readout resonators. \textbf{b)} The topological diagram of qubits and couplers. Dark gray denotes qubits, light blue denotes couplers. \textbf{c)} Simplified circuit schematic of two qubits coupled via a coupler.}
\label{fig1}
\end{figure*}

\textit{Zuchongzhi} 3.0 quantum processor marks a significant upgrade from its predecessor, \textit{Zuchongzhi} 2.0, with a notable increase in both the quantity and quality of qubits. It now houses 105 Transmon qubits, arrayed in 15 rows and 7 columns, forming a two-dimensional rectangular lattice as depicted in Fig.~\ref{fig1}. We conducted experiments with a maximum of 83 qubits selected from the processor.

One of the most significant advancements in \textit{Zuchongzhi} 3.0 quantum processor is the enhancement of coherence time. This improvement is achieved through several key strategies. First, we optimize the circuit parameters of the qubits, including the capacitance and the Josephson inductance, to reduce sensitivities to charge and flux noise. Second, we optimize the electric field distribution by modifying the shape of the qubit capacitor pads, which minimizes surface dielectric loss. Third, the attenuator configuration in the wiring is upgraded to mitigate noise from room-temperature electronics, significantly improving the dephasing time. Finally, we update the chip fabrication procedure by lithographically defining base components made of tantalum on the top sapphire substrate and aluminium on the bottom sapphire substrate, 
which are then bonded together using an indium bump flip-chip technique. This approach reduces the contamination at the interface and enhances the relaxation time of qubits. \hhl{As a result, we improve the relaxation time ($T_{1}$) to 72 $\mu s$ and the dephasing time ($T_{2,CPMG}$) to 58 $\mu s$}.

\begin{figure*}[!htbp]
\begin{center}
\includegraphics[width=1\linewidth]{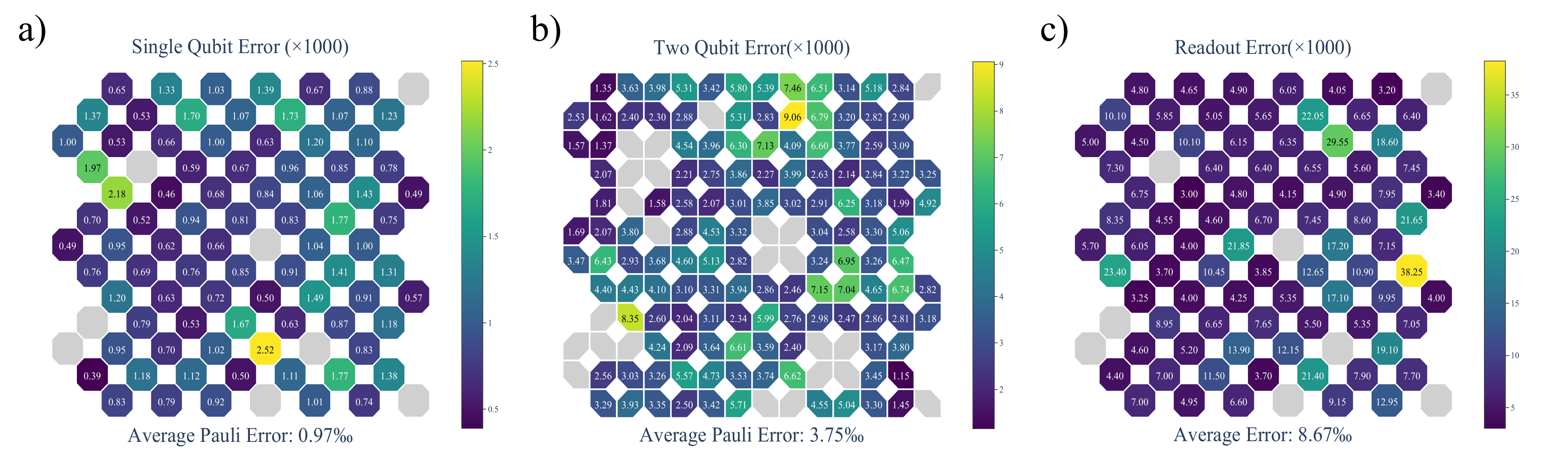}
\end{center}
\caption{\textbf{Gate and readout performance of the selected 83 qubits. }\textbf{a) }The single-qubit gate error, measured by XEB experiment, has an average value of  0.97$\permil $ and a duration of 28 ns. \textbf{b)} The two-qubit gate error used in the experiment has an average value of 3.75$\permil $ with a gate time of 45 ns. \textbf{c)} The average readout error rate is 8.67$\permil $, achieved through active reset and 0-2 state readout, which improves readout fidelity while reducing the sampling interval to 400 $\mu$s. The provided values correspond to the simultaneous operation of all selected qubits. For the detailed calibration data on the complete set of 105 qubits, refer to the Supplemental Material.}
\label{fig2}
\end{figure*}

The calibration processes for single-qubit gates and iSwap-like gates are similar to those employed in the \textit{Zuchongzhi} 2.0. Due to advancements in coherence time, the average Pauli error for  single-qubit gates($e_{1}$) and iSwap-like gates($e_{2}$) has been reduced to $0.10 \%$ and $0.38 \%$ respectively (as depicted in Fig.~\ref{fig2} (a) and (b)), when all gates are applied simultaneously.

The performance of readout is another significant advancement in the  \textit{Zuchongzhi} 3.0. To achieve fast readout with high fidelity, we increased the coupling strength between qubits and readout resonators to approximately 130 MHz and tuned the linewidths of the readout resonators to about 10 MHz. However, the increased coupling strength and linewidth result in a decrease in relaxation time. To address this, we optimized the design of the bandpass filter for dispersive qubit measurement, protecting the qubit from the Purcell effect. 

Additionally, before each sampling task, we perform three rounds of measurement and apply the corresponding single-qubit gate to reset the qubit to the state $\ket{0}$. This method reduces the impact of thermal noise on state preparation and shortens the duration of each sampling. After these optimizations, the average readout error across 83 qubits has been suppressed to 0.82\% (as depicted in Fig.~\ref{fig2} (c)).

\begin{figure*}[!htbp]
\begin{center}
\includegraphics[width=1\linewidth]{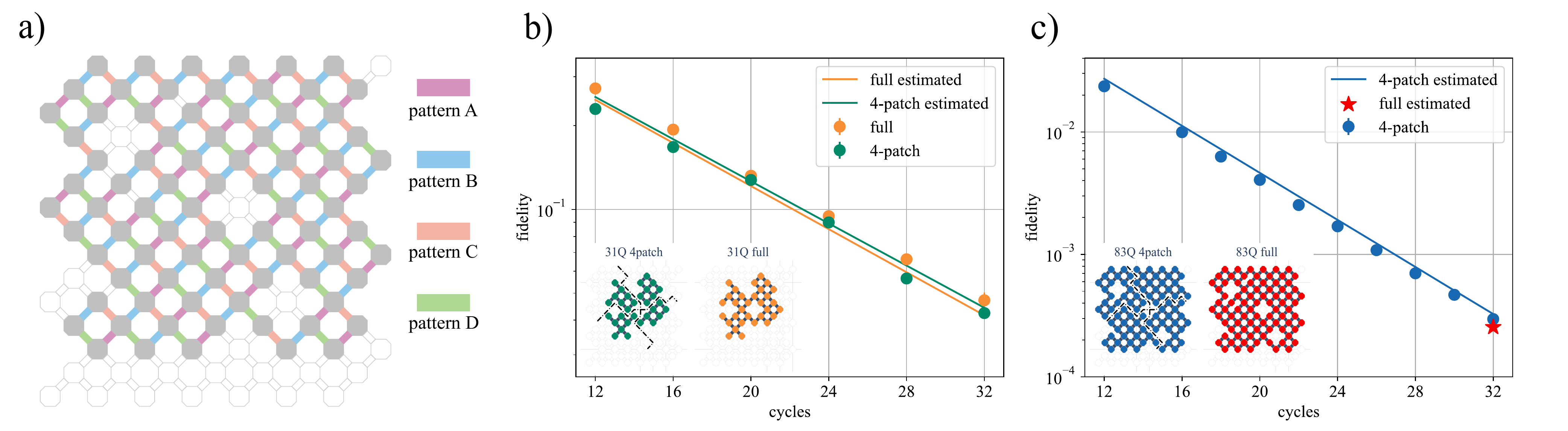}
\end{center}
\caption{\textbf{Experiment and estimate fidelity of random circuit sampling experiment for 31 qubits and 83 qubits. a)} The pattern diagram of the random circuit sampling experiment for 83 qubits. The iSWAP-like gates are selected from the patterns labeled A, B, C, and D, arranged in the sequence ABCDCDBA. The grey octagons denote functional qubits while purple, blue, orange, and green lines represent the iSWAP-like gates associated with the four pattern A,B,C,D respectively. Additionally, discarded qubits and couplers are indicated by empty octagons and lines. \textbf{b)} The green and blue dots respectively represent the experimental values of the 4-patch circuits and full circuits with 31 qubits over 12-36 cycles. The corresponding solid lines denote the estimated values for these circuits. The inserted topological diagram illustrates the specific configuration of 31 qubits. \textbf{c)} 
The blue dots and line correspond to the experimental and estimated values, respectively, of the 83-qubit 4-patch circuit. The red five-pointed star signifies the estimated value of the 83-qubit full circuit, where \hhl{410} million bitstrings are sampled. The inserted topological diagram depicts the specific configuration of 83 qubits.}
\label{fig3}
\end{figure*}

\section{Large-Scale Random Circuit Sampling}
After the initial calibration, we proceed with random quantum circuit sampling to evaluate the overall performance of the quantum processor. The random quantum circuit is designed in accordance with the method outlined in Ref.~\cite{huang2024design} to widen the performance gap between quantum computing and classical simulation. Notably, the two-qubit iSWAP-like gates within each layer of two-qubit gates are applied following a specific pattern, denoted by A, B, C, and D, as illustrated in Fig.~\ref{fig3} (a), and are executed in the sequence of ABCD-CDAB in each cycle. The single-qubit gates in each cycle are selected at random from the set $\{ \sqrt{X}, \sqrt{Y}, \sqrt{W} \}$.

Verifying the fidelity of the full random quantum circuit is challenging due to the inability to simulate its ideal output classically. To address this, patch circuits are utilized for the verification of large-scale random quantum circuits. These patch circuits are crafted by selectively removing a portion of the two-qubit gates between the patches. The entire circuit can be divided into two independent segments, termed as 2-patch, or into four segments, known as 4-patch. The more divisions made, the more feasible the simulation becomes; however, the anticipated fidelity slightly increases due to the reduction in the number of two-qubit gates executed. We implement 2-patch, 4-patch, and the full version of the circuits, scaling from \hhl{12 to 32 cycles with 31 qubits each}, and compute the linear XEB fidelities $F_{\text{XEB}}$ for the respective output bitstrings. The experimental results, as detailed in Fig.~\ref{fig3}(b) (the results of 2-patch are displayed in \hhl{Fig. S7 in Supplemental Material}), reveal that the average fidelity ratios of the 4-patch circuit to the full circuit fidelity is \hhl{1.05}. 
This high degree of correspondence indicates the effectiveness of the verification circuits in ensuring the fidelity of quantum computations.

\begin{table*}[!htbp]
\centering
\caption{\textbf{Estimated classical computational cost for different experiments.} The Frontier supercomputer boasts a theoretical peak performance of $1.685 \times 10^{18}$ FLOPS. In our estimations, we presume a 20\% FLOP efficiency and convert the machine FLOPS to single-precision complex FLOPS. We provide two scenarios: one with 9.2 PB of memory (the actual memory of Frontier) and another with 762.2 PB (combining Frontier's actual memory with all storage, which is an impractical situation).}
\renewcommand\arraystretch{1.2}
\begin{tabular}{|c|c|ccc|ccc|}
\hline
\multirow{2}{*}{\textbf{Experiment}}
& \multirow{2}{*}{\textbf{Fidelity}}
& \multicolumn{3}{c|}{\textbf{Memory Constraint: 9.2PB}}                            & \multicolumn{3}{c|}{\textbf{Memory Constraint: 762.2PB}}  \\ \cline{3-8}
& 
& \multicolumn{1}{c|}{\makecell{1 Amplitude \\ (FLOP)}}
& \multicolumn{1}{c|}{\makecell{1 Million Noisy \\ Samples (FLOP)}}
& \makecell{Runtime on \\  Frontier} 
& \multicolumn{1}{c|}{\makecell{1 Amplitude \\ (FLOP)}}        
& \multicolumn{1}{c|}{\makecell{1 Million Noisy \\ Samples (FLOP)}}
& \makecell{Runtime on \\ Frontier}  \\ \hline\hline
Sycamore-53-20   & $2.2 \times 10^{-3}$ & \multicolumn{1}{c|}{$7.2 \times 10^{18}$} & \multicolumn{1}{c|}{$6.5 \times 10^{16}$} & 1.6 s      & \multicolumn{1}{c|}{$5.9 \times 10^{18}$} & \multicolumn{1}{c|}{$6.1 \times 10^{16}$} & 1.5 s             \\ \hline
\textit{Zuchongzhi}-56-20 & $6.6 \times 10^{-4}$ & \multicolumn{1}{c|}{$9.3 \times 10^{19}$} & \multicolumn{1}{c|}{$2.2 \times 10^{17}$} & 5.3 s      & \multicolumn{1}{c|}{$1.0 \times 10^{20}$} & \multicolumn{1}{c|}{$1.5 \times 10^{17}$} & 3.6 s              \\ \hline
\textit{Zuchongzhi}-60-24 & $3.7 \times 10^{-4}$  & \multicolumn{1}{c|}{$3.2 \times 10^{21}$} & \multicolumn{1}{c|}{$1.6 \times 10^{19}$} & 384.0 s      & \multicolumn{1}{c|}{$3.0 \times 10^{21}$} & \multicolumn{1}{c|}{$2.3 \times 10^{18}$} & 55.2 s             \\ \hline
Sycamore-70-24    & $1.7 \times 10^{-3}$ & \multicolumn{1}{c|}{$1.7 \times 10^{25}$} & \multicolumn{1}{c|}{$8.2 \times 10^{25}$} & 62.1 yr      & \multicolumn{1}{c|}{$3.2 \times 10^{24}$} & \multicolumn{1}{c|}{$1.4 \times 10^{24}$} & 1.1 yr            \\ \hline
Sycamore-67-32   & $1.5 \times 10^{-3}$ & \multicolumn{1}{c|}{$8.2 \times 10^{28}$} & \multicolumn{1}{c|}{$4.7 \times 10^{27}$} & $3.6\times 10^3$ yr      & \multicolumn{1}{c|}{$1.3 \times 10^{26}$} & \multicolumn{1}{c|}{$9.6 \times 10^{24}$} & $7.2$ yr              \\ \hline
\textit{Zuchongzhi-83-3}2 & $2.5 \times 10^{-4}$ & \multicolumn{1}{c|}{$5.1 \times 10^{31}$} & \multicolumn{1}{c|}{$8.4 \times 10^{33}$} & $6.4\times 10^{9}$ yr      & \multicolumn{1}{c|}{$1.3 \times 10^{29}$} & \multicolumn{1}{c|}{$7.5 \times 10^{31}$} & $5.7 \times 10^7$ yr            \\ \hline
\end{tabular}
\end{table*}

Such an outstanding quantum processor allows us to run random circuit sampling on a larger scale than before. As shown in Fig.~\ref{fig3}(c), we have achieved random circuit sampling of \hhl{83-qubit circuits with 12 to 32 cycles}. 
For the largest full circuit featuring \hhl{83 qubits and 32 cycles}, we have collected a total of approximately \hhl{$4.1 \times 10^8$} bitstrings. To assess its fidelity, we also gathered corresponding bitstrings from 4-patch circuits, which exhibit an experimental fidelity of \hhl{0.030\%}, while the estimated fidelity stands at 0.033\%. This high degree of correspondence indicates that, even at a large scale of qubits and high circuit depth, employing the discrete error model to estimate fidelity remains highly reliable. Consequently, we can estimate the fidelity of the full circuit with \hhl{83 qubits and 32 cycles} to be \hhl{0.025\%}.

\section{Computational Cost Estimation}

The current cutting-edge classical algorithm for simulating random quantum circuits is the tensor network algorithm~\cite{MarkovShi2008,GuoWu2019,VillalongaMandra2019,villalonga2020establishing,HuangChen2020,pan2021simulating,guo2021verifying,pan2022solving,Fu2024surpassing,zhao2024leapfrogging}. We employ this method to evaluate the classical computational cost of our harderst circuit, featuring 80 qubits and 32 cycles. Considering memory constraints, we have examined the following two scenarios using the state of the art method~\cite{Fu2024surpassing,zhao2024leapfrogging}.

The first scenario involves capping the memory at 9.2 petabytes (PB), which is the memory size of the current most powerful supercomputer, Frontier. The estimated number of floating-point operations required to generate a million uncorrelated bitstrings with a fidelity of 0.025\% from an 83-qubit, 32-cycle random circuit using a classical computer is \(8.4 \times 10^{33}\). In contrast, the latest quantum computational advantage experiment by Google~\cite{morvan2023phase}, SYC-67, has an estimated classical simulation complexity of $4.7 \times 10^{27}$ for replicating the same number of bitstrings with fidelity that matches its experiment. Hence, the classical cost of simulating our most challenging random quantum circuit is six orders of magnitude higher than that of SYC-67. For our estimates, we utilize the specifications of the Frontier supercomputer, which boasts a theoretical peak performance of \(1.685 \times 10^{18}\) single precision FLOPs per second. We assume a 20\% FLOP efficiency and take into account the low target fidelity of the simulation in the computational cost. Each single precision complex FLOP necessitates 8 machine FLOPs. Under these conditions, the projected time for classical simulation of our most challenging random quantum circuit is \(6.4 \times 10^{9}\) years using the current most powerful supercomputer, Frontier.

The complexity of tensor network algorithms is influenced by memory limitations, and we have further contemplated the scenario of virtually unlimited memory as an estimated lower bound for the sampling cost, although this situation is already unrealistic. By setting the memory limit to over 762.2 PB (considering both memory and total storage of Frontier as part of the memory), we estimate the number of floating-point operations needed to generate a million uncorrelated bitstrings of the same fidelity from our most challenging 80-qubit, 32-cycle random circuit remains high, at \(7.5 \times 10^{31}\). Consequently, the estimated classical simulation time is an immense \(5.7 \times 10^{7}\) years, which underscores the robustness of our quantum advantage.


\begin{figure}[!htbp]
\begin{center}
\includegraphics[width=1\linewidth]{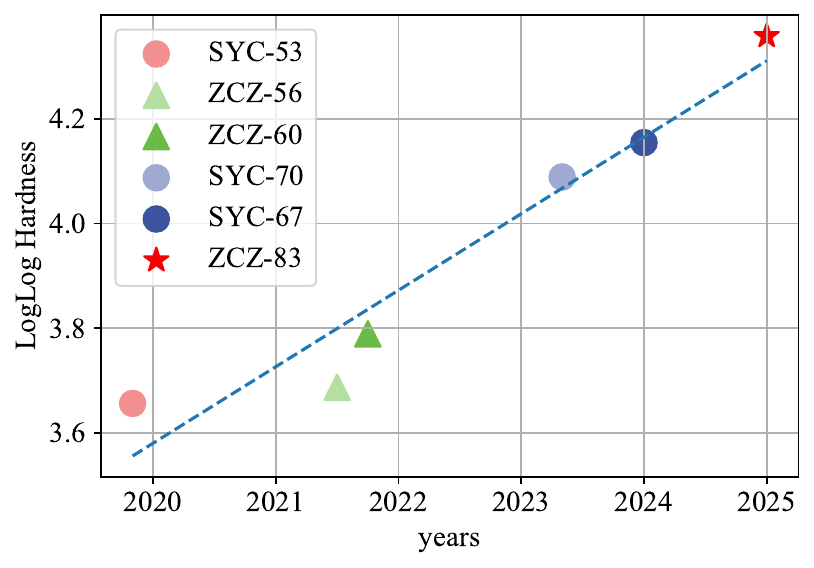}
\end{center}
\caption{\textbf{Progress on random circuit sampling.} The evolution of the time complexity in random circuit sampling experiments. The dotted line illustrates the pattern of doubly-exponential growth. SYC and ZCZ respectively denote the Sycamore and \textit{Zuchongzhi} processors.}
\label{fig4}
\end{figure}


\section{Conclusion} \label{sec:summary}

The \textit{Zuchongzhi} 3.0, an advanced superconducting quantum computer prototype with its 105 qubits and exceptional operational fidelities, not only ups the ante in terms of the number of qubits but also enhances the precision of quantum manipulation. This dual advancement is key to expanding our quantum computing capabilities. Based on this robust platform, we have successfully executed a larger scale random circuit sampling than previously achieved by Google~\cite{morvan2023phase}, further widening the gap in computational capabilities between classical and quantum computing. Our work advances the discourse on quantum computing by providing empirical evidence of the technology's potential to revolutionize computational tasks. It serves as both a testament to the progress in quantum hardware and a foundation for practical applications. Scaling up in qubits and circuit complexity enhances our capacity to address sophisticated challenges in optimization~\cite{guan2024single}, machine learning~\cite{huang2021experimental,liu2021hybrid,gong2023quantum}, and drug discovery.


\begin{acknowledgments}

\textbf{Funding:}
This research was supported by the Innovation Program for Quantum Science and Technology (Grant No.~2021ZD0300200), Anhui Initiative in Quantum Information Technologies, the Special funds from Jinan Science and Technology Bureau and Jinan high tech Zone Management Committee, Shanghai Municipal Science and Technology Major Project (Grant No.~2019SHZDZX01), and the New Cornerstone Science Foundation through the XPLORER PRIZE. M. Gong was sponsored by National Natural Science Foundation of China (Grants No.~T2322024), Shanghai Rising-Star Program (Grant No.~23QA1410000) and the Youth Innovation Promotion Association of CAS (Grant No.~2022460). H.Huang was supported from the National Natural Science Foundation of China (Grant No. 12274464), and Natural Science Foundation of Henan (Grant No. 242300421049). C.Zha was sponsored by Shanghai Science and Technology Development Funds (Grant No.~23YF1452500) and the
 China Postdoctoral Science Foundation (Grant No.~2023M740964). J.Li was supported by the Key-Area Research
 and Development Program of Guangdong Province (2020B0303060001).
\textbf{Author contributions:} X.Z., and J.-W.P. conceived the research. Y.Z., X.Z., and H.H. provided the theoretical calculations. D.G., D.F., and C.Z. designed and perform the experiment. C.P. conceived the electronics and software. All authors contributed to the discussion of the results and the development of this work. X.Z. and J.-W.P. supervised the whole project.

\end{acknowledgments}

\bibliographystyle{apsrev4-1}
\bibliography{references}

\begin{thebibliography}{31}%
\makeatletter
\providecommand \@ifxundefined [1]{%
 \@ifx{#1\undefined}
}%
\providecommand \@ifnum [1]{%
 \ifnum #1\expandafter \@firstoftwo
 \else \expandafter \@secondoftwo
 \fi
}%
\providecommand \@ifx [1]{%
 \ifx #1\expandafter \@firstoftwo
 \else \expandafter \@secondoftwo
 \fi
}%
\providecommand \natexlab [1]{#1}%
\providecommand \enquote  [1]{``#1''}%
\providecommand \bibnamefont  [1]{#1}%
\providecommand \bibfnamefont [1]{#1}%
\providecommand \citenamefont [1]{#1}%
\providecommand \href@noop [0]{\@secondoftwo}%
\providecommand \href [0]{\begingroup \@sanitize@url \@href}%
\providecommand \@href[1]{\@@startlink{#1}\@@href}%
\providecommand \@@href[1]{\endgroup#1\@@endlink}%
\providecommand \@sanitize@url [0]{\catcode `\\12\catcode `\$12\catcode `\&12\catcode `\#12\catcode `\^12\catcode `\_12\catcode `\%12\relax}%
\providecommand \@@startlink[1]{}%
\providecommand \@@endlink[0]{}%
\providecommand \url  [0]{\begingroup\@sanitize@url \@url }%
\providecommand \@url [1]{\endgroup\@href {#1}{\urlprefix }}%
\providecommand \urlprefix  [0]{URL }%
\providecommand \Eprint [0]{\href }%
\providecommand \doibase [0]{http://dx.doi.org/}%
\providecommand \selectlanguage [0]{\@gobble}%
\providecommand \bibinfo  [0]{\@secondoftwo}%
\providecommand \bibfield  [0]{\@secondoftwo}%
\providecommand \translation [1]{[#1]}%
\providecommand \BibitemOpen [0]{}%
\providecommand \bibitemStop [0]{}%
\providecommand \bibitemNoStop [0]{.\EOS\space}%
\providecommand \EOS [0]{\spacefactor3000\relax}%
\providecommand \BibitemShut  [1]{\csname bibitem#1\endcsname}%
\let\auto@bib@innerbib\@empty
\bibitem [{\citenamefont {Preskill}(2018)}]{preskill2018quantum}%
  \BibitemOpen
  \bibfield  {author} {\bibinfo {author} {\bibfnamefont {J.}~\bibnamefont {Preskill}},\ }\href@noop {} {\bibfield  {journal} {\bibinfo  {journal} {Quantum}\ }\textbf {\bibinfo {volume} {2}},\ \bibinfo {pages} {79} (\bibinfo {year} {2018})}\BibitemShut {NoStop}%
\bibitem [{\citenamefont {Wu}\ \emph {et~al.}(2021)\citenamefont {Wu}, \citenamefont {Bao}, \citenamefont {Cao}, \citenamefont {Chen}, \citenamefont {Chen}, \citenamefont {Chen}, \citenamefont {Chung}, \citenamefont {Deng}, \citenamefont {Du}, \citenamefont {Fan} \emph {et~al.}}]{Wu2021strong}%
  \BibitemOpen
  \bibfield  {author} {\bibinfo {author} {\bibfnamefont {Y.}~\bibnamefont {Wu}}, \bibinfo {author} {\bibfnamefont {W.-S.}\ \bibnamefont {Bao}}, \bibinfo {author} {\bibfnamefont {S.}~\bibnamefont {Cao}}, \bibinfo {author} {\bibfnamefont {F.}~\bibnamefont {Chen}}, \bibinfo {author} {\bibfnamefont {M.-C.}\ \bibnamefont {Chen}}, \bibinfo {author} {\bibfnamefont {X.}~\bibnamefont {Chen}}, \bibinfo {author} {\bibfnamefont {T.-H.}\ \bibnamefont {Chung}}, \bibinfo {author} {\bibfnamefont {H.}~\bibnamefont {Deng}}, \bibinfo {author} {\bibfnamefont {Y.}~\bibnamefont {Du}}, \bibinfo {author} {\bibfnamefont {D.}~\bibnamefont {Fan}},  \emph {et~al.},\ }\href@noop {} {\bibfield  {journal} {\bibinfo  {journal} {Physical Review Letters}\ }\textbf {\bibinfo {volume} {127}},\ \bibinfo {pages} {180501} (\bibinfo {year} {2021})}\BibitemShut {NoStop}%
\bibitem [{\citenamefont {Zhu}\ \emph {et~al.}(2022)\citenamefont {Zhu}, \citenamefont {Cao}, \citenamefont {Chen}, \citenamefont {Chen}, \citenamefont {Chen}, \citenamefont {Chung}, \citenamefont {Deng}, \citenamefont {Du}, \citenamefont {Fan}, \citenamefont {Gong} \emph {et~al.}}]{zhu2022quantum}%
  \BibitemOpen
  \bibfield  {author} {\bibinfo {author} {\bibfnamefont {Q.}~\bibnamefont {Zhu}}, \bibinfo {author} {\bibfnamefont {S.}~\bibnamefont {Cao}}, \bibinfo {author} {\bibfnamefont {F.}~\bibnamefont {Chen}}, \bibinfo {author} {\bibfnamefont {M.-C.}\ \bibnamefont {Chen}}, \bibinfo {author} {\bibfnamefont {X.}~\bibnamefont {Chen}}, \bibinfo {author} {\bibfnamefont {T.-H.}\ \bibnamefont {Chung}}, \bibinfo {author} {\bibfnamefont {H.}~\bibnamefont {Deng}}, \bibinfo {author} {\bibfnamefont {Y.}~\bibnamefont {Du}}, \bibinfo {author} {\bibfnamefont {D.}~\bibnamefont {Fan}}, \bibinfo {author} {\bibfnamefont {M.}~\bibnamefont {Gong}},  \emph {et~al.},\ }\href@noop {} {\bibfield  {journal} {\bibinfo  {journal} {Science Bulletin}\ }\textbf {\bibinfo {volume} {67}},\ \bibinfo {pages} {240} (\bibinfo {year} {2022})}\BibitemShut {NoStop}%
\bibitem [{\citenamefont {Arute}\ \emph {et~al.}(2019)\citenamefont {Arute}, \citenamefont {Arya}, \citenamefont {Babbush}, \citenamefont {Bacon}, \citenamefont {Bardin}, \citenamefont {Barends}, \citenamefont {Biswas}, \citenamefont {Boixo}, \citenamefont {Brandao}, \citenamefont {Buell} \emph {et~al.}}]{arute2019quantum}%
  \BibitemOpen
  \bibfield  {author} {\bibinfo {author} {\bibfnamefont {F.}~\bibnamefont {Arute}}, \bibinfo {author} {\bibfnamefont {K.}~\bibnamefont {Arya}}, \bibinfo {author} {\bibfnamefont {R.}~\bibnamefont {Babbush}}, \bibinfo {author} {\bibfnamefont {D.}~\bibnamefont {Bacon}}, \bibinfo {author} {\bibfnamefont {J.~C.}\ \bibnamefont {Bardin}}, \bibinfo {author} {\bibfnamefont {R.}~\bibnamefont {Barends}}, \bibinfo {author} {\bibfnamefont {R.}~\bibnamefont {Biswas}}, \bibinfo {author} {\bibfnamefont {S.}~\bibnamefont {Boixo}}, \bibinfo {author} {\bibfnamefont {F.~G.}\ \bibnamefont {Brandao}}, \bibinfo {author} {\bibfnamefont {D.~A.}\ \bibnamefont {Buell}},  \emph {et~al.},\ }\href@noop {} {\bibfield  {journal} {\bibinfo  {journal} {Nature}\ }\textbf {\bibinfo {volume} {574}},\ \bibinfo {pages} {505} (\bibinfo {year} {2019})}\BibitemShut {NoStop}%
\bibitem [{\citenamefont {Morvan}\ \emph {et~al.}(2024)\citenamefont {Morvan}, \citenamefont {Villalonga}, \citenamefont {Mi}, \citenamefont {Mandra}, \citenamefont {Bengtsson}, \citenamefont {Klimov}, \citenamefont {Chen}, \citenamefont {Hong}, \citenamefont {Erickson}, \citenamefont {Drozdov} \emph {et~al.}}]{morvan2023phase}%
  \BibitemOpen
  \bibfield  {author} {\bibinfo {author} {\bibfnamefont {A.}~\bibnamefont {Morvan}}, \bibinfo {author} {\bibfnamefont {B.}~\bibnamefont {Villalonga}}, \bibinfo {author} {\bibfnamefont {X.}~\bibnamefont {Mi}}, \bibinfo {author} {\bibfnamefont {S.}~\bibnamefont {Mandra}}, \bibinfo {author} {\bibfnamefont {A.}~\bibnamefont {Bengtsson}}, \bibinfo {author} {\bibfnamefont {P.}~\bibnamefont {Klimov}}, \bibinfo {author} {\bibfnamefont {Z.}~\bibnamefont {Chen}}, \bibinfo {author} {\bibfnamefont {S.}~\bibnamefont {Hong}}, \bibinfo {author} {\bibfnamefont {C.}~\bibnamefont {Erickson}}, \bibinfo {author} {\bibfnamefont {I.}~\bibnamefont {Drozdov}},  \emph {et~al.},\ }\href@noop {} {\bibfield  {journal} {\bibinfo  {journal} {Nature}\ }\textbf {\bibinfo {volume} {634}},\ \bibinfo {pages} {328} (\bibinfo {year} {2024})}\BibitemShut {NoStop}%
\bibitem [{\citenamefont {Zhong}\ \emph {et~al.}(2020)\citenamefont {Zhong}, \citenamefont {Wang}, \citenamefont {Deng}, \citenamefont {Chen}, \citenamefont {Peng}, \citenamefont {Luo}, \citenamefont {Qin}, \citenamefont {Wu}, \citenamefont {Ding}, \citenamefont {Hu} \emph {et~al.}}]{zhong2020quantum}%
  \BibitemOpen
  \bibfield  {author} {\bibinfo {author} {\bibfnamefont {H.-S.}\ \bibnamefont {Zhong}}, \bibinfo {author} {\bibfnamefont {H.}~\bibnamefont {Wang}}, \bibinfo {author} {\bibfnamefont {Y.-H.}\ \bibnamefont {Deng}}, \bibinfo {author} {\bibfnamefont {M.-C.}\ \bibnamefont {Chen}}, \bibinfo {author} {\bibfnamefont {L.-C.}\ \bibnamefont {Peng}}, \bibinfo {author} {\bibfnamefont {Y.-H.}\ \bibnamefont {Luo}}, \bibinfo {author} {\bibfnamefont {J.}~\bibnamefont {Qin}}, \bibinfo {author} {\bibfnamefont {D.}~\bibnamefont {Wu}}, \bibinfo {author} {\bibfnamefont {X.}~\bibnamefont {Ding}}, \bibinfo {author} {\bibfnamefont {Y.}~\bibnamefont {Hu}},  \emph {et~al.},\ }\href@noop {} {\bibfield  {journal} {\bibinfo  {journal} {Science}\ }\textbf {\bibinfo {volume} {370}},\ \bibinfo {pages} {1460} (\bibinfo {year} {2020})}\BibitemShut {NoStop}%
\bibitem [{\citenamefont {Deng}\ \emph {et~al.}(2023)\citenamefont {Deng}, \citenamefont {Gu}, \citenamefont {Liu}, \citenamefont {Gong}, \citenamefont {Su}, \citenamefont {Zhang}, \citenamefont {Tang}, \citenamefont {Jia}, \citenamefont {Xu}, \citenamefont {Chen} \emph {et~al.}}]{deng2023gaussian}%
  \BibitemOpen
  \bibfield  {author} {\bibinfo {author} {\bibfnamefont {Y.-H.}\ \bibnamefont {Deng}}, \bibinfo {author} {\bibfnamefont {Y.-C.}\ \bibnamefont {Gu}}, \bibinfo {author} {\bibfnamefont {H.-L.}\ \bibnamefont {Liu}}, \bibinfo {author} {\bibfnamefont {S.-Q.}\ \bibnamefont {Gong}}, \bibinfo {author} {\bibfnamefont {H.}~\bibnamefont {Su}}, \bibinfo {author} {\bibfnamefont {Z.-J.}\ \bibnamefont {Zhang}}, \bibinfo {author} {\bibfnamefont {H.-Y.}\ \bibnamefont {Tang}}, \bibinfo {author} {\bibfnamefont {M.-H.}\ \bibnamefont {Jia}}, \bibinfo {author} {\bibfnamefont {J.-M.}\ \bibnamefont {Xu}}, \bibinfo {author} {\bibfnamefont {M.-C.}\ \bibnamefont {Chen}},  \emph {et~al.},\ }\href@noop {} {\bibfield  {journal} {\bibinfo  {journal} {Physical Review Letters}\ }\textbf {\bibinfo {volume} {131}},\ \bibinfo {pages} {150601} (\bibinfo {year} {2023})}\BibitemShut {NoStop}%
\bibitem [{\citenamefont {Zhong}\ \emph {et~al.}(2021)\citenamefont {Zhong}, \citenamefont {Deng}, \citenamefont {Qin}, \citenamefont {Wang}, \citenamefont {Chen}, \citenamefont {Peng}, \citenamefont {Luo}, \citenamefont {Wu}, \citenamefont {Gong}, \citenamefont {Su} \emph {et~al.}}]{zhong2021phase}%
  \BibitemOpen
  \bibfield  {author} {\bibinfo {author} {\bibfnamefont {H.-S.}\ \bibnamefont {Zhong}}, \bibinfo {author} {\bibfnamefont {Y.-H.}\ \bibnamefont {Deng}}, \bibinfo {author} {\bibfnamefont {J.}~\bibnamefont {Qin}}, \bibinfo {author} {\bibfnamefont {H.}~\bibnamefont {Wang}}, \bibinfo {author} {\bibfnamefont {M.-C.}\ \bibnamefont {Chen}}, \bibinfo {author} {\bibfnamefont {L.-C.}\ \bibnamefont {Peng}}, \bibinfo {author} {\bibfnamefont {Y.-H.}\ \bibnamefont {Luo}}, \bibinfo {author} {\bibfnamefont {D.}~\bibnamefont {Wu}}, \bibinfo {author} {\bibfnamefont {S.-Q.}\ \bibnamefont {Gong}}, \bibinfo {author} {\bibfnamefont {H.}~\bibnamefont {Su}},  \emph {et~al.},\ }\href@noop {} {\bibfield  {journal} {\bibinfo  {journal} {Physical Review Letters}\ }\textbf {\bibinfo {volume} {127}},\ \bibinfo {pages} {180502} (\bibinfo {year} {2021})}\BibitemShut {NoStop}%
\bibitem [{\citenamefont {Madsen}\ \emph {et~al.}(2022)\citenamefont {Madsen}, \citenamefont {Laudenbach}, \citenamefont {Askarani}, \citenamefont {Rortais}, \citenamefont {Vincent}, \citenamefont {Bulmer}, \citenamefont {Miatto}, \citenamefont {Neuhaus}, \citenamefont {Helt}, \citenamefont {Collins} \emph {et~al.}}]{madsen2022quantum}%
  \BibitemOpen
  \bibfield  {author} {\bibinfo {author} {\bibfnamefont {L.~S.}\ \bibnamefont {Madsen}}, \bibinfo {author} {\bibfnamefont {F.}~\bibnamefont {Laudenbach}}, \bibinfo {author} {\bibfnamefont {M.~F.}\ \bibnamefont {Askarani}}, \bibinfo {author} {\bibfnamefont {F.}~\bibnamefont {Rortais}}, \bibinfo {author} {\bibfnamefont {T.}~\bibnamefont {Vincent}}, \bibinfo {author} {\bibfnamefont {J.~F.}\ \bibnamefont {Bulmer}}, \bibinfo {author} {\bibfnamefont {F.~M.}\ \bibnamefont {Miatto}}, \bibinfo {author} {\bibfnamefont {L.}~\bibnamefont {Neuhaus}}, \bibinfo {author} {\bibfnamefont {L.~G.}\ \bibnamefont {Helt}}, \bibinfo {author} {\bibfnamefont {M.~J.}\ \bibnamefont {Collins}},  \emph {et~al.},\ }\href@noop {} {\bibfield  {journal} {\bibinfo  {journal} {Nature}\ }\textbf {\bibinfo {volume} {606}},\ \bibinfo {pages} {75} (\bibinfo {year} {2022})}\BibitemShut {NoStop}%
\bibitem [{\citenamefont {Boixo}\ \emph {et~al.}(2018)\citenamefont {Boixo}, \citenamefont {Isakov}, \citenamefont {Smelyanskiy}, \citenamefont {Babbush}, \citenamefont {Ding}, \citenamefont {Jiang}, \citenamefont {Bremner}, \citenamefont {Martinis},\ and\ \citenamefont {Neven}}]{boixo2018characterizing}%
  \BibitemOpen
  \bibfield  {author} {\bibinfo {author} {\bibfnamefont {S.}~\bibnamefont {Boixo}}, \bibinfo {author} {\bibfnamefont {S.~V.}\ \bibnamefont {Isakov}}, \bibinfo {author} {\bibfnamefont {V.~N.}\ \bibnamefont {Smelyanskiy}}, \bibinfo {author} {\bibfnamefont {R.}~\bibnamefont {Babbush}}, \bibinfo {author} {\bibfnamefont {N.}~\bibnamefont {Ding}}, \bibinfo {author} {\bibfnamefont {Z.}~\bibnamefont {Jiang}}, \bibinfo {author} {\bibfnamefont {M.~J.}\ \bibnamefont {Bremner}}, \bibinfo {author} {\bibfnamefont {J.~M.}\ \bibnamefont {Martinis}}, \ and\ \bibinfo {author} {\bibfnamefont {H.}~\bibnamefont {Neven}},\ }\href@noop {} {\bibfield  {journal} {\bibinfo  {journal} {Nature Physics}\ }\textbf {\bibinfo {volume} {14}},\ \bibinfo {pages} {595} (\bibinfo {year} {2018})}\BibitemShut {NoStop}%
\bibitem [{\citenamefont {Bouland}\ \emph {et~al.}(2019)\citenamefont {Bouland}, \citenamefont {Fefferman}, \citenamefont {Nirkhe},\ and\ \citenamefont {Vazirani}}]{bouland2019complexity}%
  \BibitemOpen
  \bibfield  {author} {\bibinfo {author} {\bibfnamefont {A.}~\bibnamefont {Bouland}}, \bibinfo {author} {\bibfnamefont {B.}~\bibnamefont {Fefferman}}, \bibinfo {author} {\bibfnamefont {C.}~\bibnamefont {Nirkhe}}, \ and\ \bibinfo {author} {\bibfnamefont {U.}~\bibnamefont {Vazirani}},\ }\href@noop {} {\bibfield  {journal} {\bibinfo  {journal} {Nature Physics}\ }\textbf {\bibinfo {volume} {15}},\ \bibinfo {pages} {159} (\bibinfo {year} {2019})}\BibitemShut {NoStop}%
\bibitem [{\citenamefont {Aaronson}\ and\ \citenamefont {Chen}(2016)}]{aaronson2016complexity}%
  \BibitemOpen
  \bibfield  {author} {\bibinfo {author} {\bibfnamefont {S.}~\bibnamefont {Aaronson}}\ and\ \bibinfo {author} {\bibfnamefont {L.}~\bibnamefont {Chen}},\ }\href@noop {} {\bibfield  {journal} {\bibinfo  {journal} {arXiv:1612.05903}\ } (\bibinfo {year} {2016})}\BibitemShut {NoStop}%
\bibitem [{\citenamefont {Gong}\ \emph {et~al.}(2021)\citenamefont {Gong}, \citenamefont {Wang}, \citenamefont {Zha}, \citenamefont {Chen}, \citenamefont {Huang}, \citenamefont {Wu}, \citenamefont {Zhu}, \citenamefont {Zhao}, \citenamefont {Li}, \citenamefont {Guo} \emph {et~al.}}]{Gong2021quantum}%
  \BibitemOpen
  \bibfield  {author} {\bibinfo {author} {\bibfnamefont {M.}~\bibnamefont {Gong}}, \bibinfo {author} {\bibfnamefont {S.}~\bibnamefont {Wang}}, \bibinfo {author} {\bibfnamefont {C.}~\bibnamefont {Zha}}, \bibinfo {author} {\bibfnamefont {M.-C.}\ \bibnamefont {Chen}}, \bibinfo {author} {\bibfnamefont {H.-L.}\ \bibnamefont {Huang}}, \bibinfo {author} {\bibfnamefont {Y.}~\bibnamefont {Wu}}, \bibinfo {author} {\bibfnamefont {Q.}~\bibnamefont {Zhu}}, \bibinfo {author} {\bibfnamefont {Y.}~\bibnamefont {Zhao}}, \bibinfo {author} {\bibfnamefont {S.}~\bibnamefont {Li}}, \bibinfo {author} {\bibfnamefont {S.}~\bibnamefont {Guo}},  \emph {et~al.},\ }\href@noop {} {\bibfield  {journal} {\bibinfo  {journal} {Science}\ }\textbf {\bibinfo {volume} {372}},\ \bibinfo {pages} {948} (\bibinfo {year} {2021})}\BibitemShut {NoStop}%
\bibitem [{\citenamefont {Cao}\ \emph {et~al.}(2023)\citenamefont {Cao}, \citenamefont {Wu}, \citenamefont {Chen}, \citenamefont {Gong}, \citenamefont {Wu}, \citenamefont {Ye}, \citenamefont {Zha}, \citenamefont {Qian}, \citenamefont {Ying}, \citenamefont {Guo} \emph {et~al.}}]{cao2023generation}%
  \BibitemOpen
  \bibfield  {author} {\bibinfo {author} {\bibfnamefont {S.}~\bibnamefont {Cao}}, \bibinfo {author} {\bibfnamefont {B.}~\bibnamefont {Wu}}, \bibinfo {author} {\bibfnamefont {F.}~\bibnamefont {Chen}}, \bibinfo {author} {\bibfnamefont {M.}~\bibnamefont {Gong}}, \bibinfo {author} {\bibfnamefont {Y.}~\bibnamefont {Wu}}, \bibinfo {author} {\bibfnamefont {Y.}~\bibnamefont {Ye}}, \bibinfo {author} {\bibfnamefont {C.}~\bibnamefont {Zha}}, \bibinfo {author} {\bibfnamefont {H.}~\bibnamefont {Qian}}, \bibinfo {author} {\bibfnamefont {C.}~\bibnamefont {Ying}}, \bibinfo {author} {\bibfnamefont {S.}~\bibnamefont {Guo}},  \emph {et~al.},\ }\href@noop {} {\bibfield  {journal} {\bibinfo  {journal} {Nature}\ }\textbf {\bibinfo {volume} {619}},\ \bibinfo {pages} {738} (\bibinfo {year} {2023})}\BibitemShut {NoStop}%
\bibitem [{\citenamefont {Zhao}\ \emph {et~al.}(2022)\citenamefont {Zhao}, \citenamefont {Ye}, \citenamefont {Huang}, \citenamefont {Zhang}, \citenamefont {Wu}, \citenamefont {Guan}, \citenamefont {Zhu}, \citenamefont {Wei}, \citenamefont {He}, \citenamefont {Cao} \emph {et~al.}}]{zhao2022realization}%
  \BibitemOpen
  \bibfield  {author} {\bibinfo {author} {\bibfnamefont {Y.}~\bibnamefont {Zhao}}, \bibinfo {author} {\bibfnamefont {Y.}~\bibnamefont {Ye}}, \bibinfo {author} {\bibfnamefont {H.-L.}\ \bibnamefont {Huang}}, \bibinfo {author} {\bibfnamefont {Y.}~\bibnamefont {Zhang}}, \bibinfo {author} {\bibfnamefont {D.}~\bibnamefont {Wu}}, \bibinfo {author} {\bibfnamefont {H.}~\bibnamefont {Guan}}, \bibinfo {author} {\bibfnamefont {Q.}~\bibnamefont {Zhu}}, \bibinfo {author} {\bibfnamefont {Z.}~\bibnamefont {Wei}}, \bibinfo {author} {\bibfnamefont {T.}~\bibnamefont {He}}, \bibinfo {author} {\bibfnamefont {S.}~\bibnamefont {Cao}},  \emph {et~al.},\ }\href@noop {} {\bibfield  {journal} {\bibinfo  {journal} {Physical Review Letters}\ }\textbf {\bibinfo {volume} {129}},\ \bibinfo {pages} {030501} (\bibinfo {year} {2022})}\BibitemShut {NoStop}%
\bibitem [{\citenamefont {Ye}\ \emph {et~al.}(2023)\citenamefont {Ye}, \citenamefont {He}, \citenamefont {Huang}, \citenamefont {Wei}, \citenamefont {Zhang}, \citenamefont {Zhao}, \citenamefont {Wu}, \citenamefont {Zhu}, \citenamefont {Guan}, \citenamefont {Cao} \emph {et~al.}}]{ye2023logical}%
  \BibitemOpen
  \bibfield  {author} {\bibinfo {author} {\bibfnamefont {Y.}~\bibnamefont {Ye}}, \bibinfo {author} {\bibfnamefont {T.}~\bibnamefont {He}}, \bibinfo {author} {\bibfnamefont {H.-L.}\ \bibnamefont {Huang}}, \bibinfo {author} {\bibfnamefont {Z.}~\bibnamefont {Wei}}, \bibinfo {author} {\bibfnamefont {Y.}~\bibnamefont {Zhang}}, \bibinfo {author} {\bibfnamefont {Y.}~\bibnamefont {Zhao}}, \bibinfo {author} {\bibfnamefont {D.}~\bibnamefont {Wu}}, \bibinfo {author} {\bibfnamefont {Q.}~\bibnamefont {Zhu}}, \bibinfo {author} {\bibfnamefont {H.}~\bibnamefont {Guan}}, \bibinfo {author} {\bibfnamefont {S.}~\bibnamefont {Cao}},  \emph {et~al.},\ }\href@noop {} {\bibfield  {journal} {\bibinfo  {journal} {Physical Review Letters}\ }\textbf {\bibinfo {volume} {131}},\ \bibinfo {pages} {210603} (\bibinfo {year} {2023})}\BibitemShut {NoStop}%
\bibitem [{\citenamefont {Huang}\ \emph {et~al.}(2024)\citenamefont {Huang}, \citenamefont {Zhao},\ and\ \citenamefont {Guo}}]{huang2024design}%
  \BibitemOpen
  \bibfield  {author} {\bibinfo {author} {\bibfnamefont {H.-L.}\ \bibnamefont {Huang}}, \bibinfo {author} {\bibfnamefont {Y.}~\bibnamefont {Zhao}}, \ and\ \bibinfo {author} {\bibfnamefont {C.}~\bibnamefont {Guo}},\ }\href@noop {} {\bibfield  {journal} {\bibinfo  {journal} {Intelligent Computing}\ }\textbf {\bibinfo {volume} {3}},\ \bibinfo {pages} {0079} (\bibinfo {year} {2024})}\BibitemShut {NoStop}%
\bibitem [{\citenamefont {Markov}\ and\ \citenamefont {Shi}(2008)}]{MarkovShi2008}%
  \BibitemOpen
  \bibfield  {author} {\bibinfo {author} {\bibfnamefont {I.~L.}\ \bibnamefont {Markov}}\ and\ \bibinfo {author} {\bibfnamefont {Y.}~\bibnamefont {Shi}},\ }\href@noop {} {\bibfield  {journal} {\bibinfo  {journal} {SIAM Journal on Computing}\ }\textbf {\bibinfo {volume} {38}},\ \bibinfo {pages} {963} (\bibinfo {year} {2008})}\BibitemShut {NoStop}%
\bibitem [{\citenamefont {Guo}\ \emph {et~al.}(2019)\citenamefont {Guo}, \citenamefont {Liu}, \citenamefont {Xiong}, \citenamefont {Xue}, \citenamefont {Fu}, \citenamefont {Huang}, \citenamefont {Qiang}, \citenamefont {Xu}, \citenamefont {Liu}, \citenamefont {Zheng} \emph {et~al.}}]{GuoWu2019}%
  \BibitemOpen
  \bibfield  {author} {\bibinfo {author} {\bibfnamefont {C.}~\bibnamefont {Guo}}, \bibinfo {author} {\bibfnamefont {Y.}~\bibnamefont {Liu}}, \bibinfo {author} {\bibfnamefont {M.}~\bibnamefont {Xiong}}, \bibinfo {author} {\bibfnamefont {S.}~\bibnamefont {Xue}}, \bibinfo {author} {\bibfnamefont {X.}~\bibnamefont {Fu}}, \bibinfo {author} {\bibfnamefont {A.}~\bibnamefont {Huang}}, \bibinfo {author} {\bibfnamefont {X.}~\bibnamefont {Qiang}}, \bibinfo {author} {\bibfnamefont {P.}~\bibnamefont {Xu}}, \bibinfo {author} {\bibfnamefont {J.}~\bibnamefont {Liu}}, \bibinfo {author} {\bibfnamefont {S.}~\bibnamefont {Zheng}},  \emph {et~al.},\ }\href@noop {} {\bibfield  {journal} {\bibinfo  {journal} {Physical Review Letters}\ }\textbf {\bibinfo {volume} {123}},\ \bibinfo {pages} {190501} (\bibinfo {year} {2019})}\BibitemShut {NoStop}%
\bibitem [{\citenamefont {Villalonga}\ \emph {et~al.}(2019)\citenamefont {Villalonga}, \citenamefont {Boixo}, \citenamefont {Nelson}, \citenamefont {Henze}, \citenamefont {Rieffel}, \citenamefont {Biswas},\ and\ \citenamefont {Mandr{\`a}}}]{VillalongaMandra2019}%
  \BibitemOpen
  \bibfield  {author} {\bibinfo {author} {\bibfnamefont {B.}~\bibnamefont {Villalonga}}, \bibinfo {author} {\bibfnamefont {S.}~\bibnamefont {Boixo}}, \bibinfo {author} {\bibfnamefont {B.}~\bibnamefont {Nelson}}, \bibinfo {author} {\bibfnamefont {C.}~\bibnamefont {Henze}}, \bibinfo {author} {\bibfnamefont {E.}~\bibnamefont {Rieffel}}, \bibinfo {author} {\bibfnamefont {R.}~\bibnamefont {Biswas}}, \ and\ \bibinfo {author} {\bibfnamefont {S.}~\bibnamefont {Mandr{\`a}}},\ }\href@noop {} {\bibfield  {journal} {\bibinfo  {journal} {npj Quantum Information}\ }\textbf {\bibinfo {volume} {5}},\ \bibinfo {pages} {1} (\bibinfo {year} {2019})}\BibitemShut {NoStop}%
\bibitem [{\citenamefont {Villalonga}\ \emph {et~al.}(2020)\citenamefont {Villalonga}, \citenamefont {Lyakh}, \citenamefont {Boixo}, \citenamefont {Neven}, \citenamefont {Humble}, \citenamefont {Biswas}, \citenamefont {Rieffel}, \citenamefont {Ho},\ and\ \citenamefont {Mandr{\`a}}}]{villalonga2020establishing}%
  \BibitemOpen
  \bibfield  {author} {\bibinfo {author} {\bibfnamefont {B.}~\bibnamefont {Villalonga}}, \bibinfo {author} {\bibfnamefont {D.}~\bibnamefont {Lyakh}}, \bibinfo {author} {\bibfnamefont {S.}~\bibnamefont {Boixo}}, \bibinfo {author} {\bibfnamefont {H.}~\bibnamefont {Neven}}, \bibinfo {author} {\bibfnamefont {T.~S.}\ \bibnamefont {Humble}}, \bibinfo {author} {\bibfnamefont {R.}~\bibnamefont {Biswas}}, \bibinfo {author} {\bibfnamefont {E.~G.}\ \bibnamefont {Rieffel}}, \bibinfo {author} {\bibfnamefont {A.}~\bibnamefont {Ho}}, \ and\ \bibinfo {author} {\bibfnamefont {S.}~\bibnamefont {Mandr{\`a}}},\ }\href@noop {} {\bibfield  {journal} {\bibinfo  {journal} {Quantum Science and Technology}\ }\textbf {\bibinfo {volume} {5}},\ \bibinfo {pages} {034003} (\bibinfo {year} {2020})}\BibitemShut {NoStop}%
\bibitem [{\citenamefont {Huang}\ \emph {et~al.}(2020)\citenamefont {Huang}, \citenamefont {Zhang}, \citenamefont {Newman}, \citenamefont {Cai}, \citenamefont {Gao}, \citenamefont {Tian}, \citenamefont {Wu}, \citenamefont {Xu}, \citenamefont {Yu}, \citenamefont {Yuan} \emph {et~al.}}]{HuangChen2020}%
  \BibitemOpen
  \bibfield  {author} {\bibinfo {author} {\bibfnamefont {C.}~\bibnamefont {Huang}}, \bibinfo {author} {\bibfnamefont {F.}~\bibnamefont {Zhang}}, \bibinfo {author} {\bibfnamefont {M.}~\bibnamefont {Newman}}, \bibinfo {author} {\bibfnamefont {J.}~\bibnamefont {Cai}}, \bibinfo {author} {\bibfnamefont {X.}~\bibnamefont {Gao}}, \bibinfo {author} {\bibfnamefont {Z.}~\bibnamefont {Tian}}, \bibinfo {author} {\bibfnamefont {J.}~\bibnamefont {Wu}}, \bibinfo {author} {\bibfnamefont {H.}~\bibnamefont {Xu}}, \bibinfo {author} {\bibfnamefont {H.}~\bibnamefont {Yu}}, \bibinfo {author} {\bibfnamefont {B.}~\bibnamefont {Yuan}},  \emph {et~al.},\ }\href@noop {} {\bibfield  {journal} {\bibinfo  {journal} {arXiv:2005.06787}\ } (\bibinfo {year} {2020})}\BibitemShut {NoStop}%
\bibitem [{\citenamefont {Pan}\ and\ \citenamefont {Zhang}(2022)}]{pan2021simulating}%
  \BibitemOpen
  \bibfield  {author} {\bibinfo {author} {\bibfnamefont {F.}~\bibnamefont {Pan}}\ and\ \bibinfo {author} {\bibfnamefont {P.}~\bibnamefont {Zhang}},\ }\href@noop {} {\bibfield  {journal} {\bibinfo  {journal} {Physical Review Letters}\ }\textbf {\bibinfo {volume} {128}},\ \bibinfo {pages} {030501} (\bibinfo {year} {2022})}\BibitemShut {NoStop}%
\bibitem [{\citenamefont {Guo}\ \emph {et~al.}(2021)\citenamefont {Guo}, \citenamefont {Zhao},\ and\ \citenamefont {Huang}}]{guo2021verifying}%
  \BibitemOpen
  \bibfield  {author} {\bibinfo {author} {\bibfnamefont {C.}~\bibnamefont {Guo}}, \bibinfo {author} {\bibfnamefont {Y.}~\bibnamefont {Zhao}}, \ and\ \bibinfo {author} {\bibfnamefont {H.-L.}\ \bibnamefont {Huang}},\ }\href@noop {} {\bibfield  {journal} {\bibinfo  {journal} {Physical Review Letters}\ }\textbf {\bibinfo {volume} {126}},\ \bibinfo {pages} {070502} (\bibinfo {year} {2021})}\BibitemShut {NoStop}%
\bibitem [{\citenamefont {Pan}\ \emph {et~al.}(2022)\citenamefont {Pan}, \citenamefont {Chen},\ and\ \citenamefont {Zhang}}]{pan2022solving}%
  \BibitemOpen
  \bibfield  {author} {\bibinfo {author} {\bibfnamefont {F.}~\bibnamefont {Pan}}, \bibinfo {author} {\bibfnamefont {K.}~\bibnamefont {Chen}}, \ and\ \bibinfo {author} {\bibfnamefont {P.}~\bibnamefont {Zhang}},\ }\href@noop {} {\bibfield  {journal} {\bibinfo  {journal} {Physical Review Letters}\ }\textbf {\bibinfo {volume} {129}},\ \bibinfo {pages} {090502} (\bibinfo {year} {2022})}\BibitemShut {NoStop}%
\bibitem [{\citenamefont {Fu}\ \emph {et~al.}(2024)\citenamefont {Fu}, \citenamefont {Su}, \citenamefont {Zhong}, \citenamefont {Zhao}, \citenamefont {Zhang}, \citenamefont {Pan}, \citenamefont {Zhang}, \citenamefont {Zhao}, \citenamefont {Chen}, \citenamefont {Lu} \emph {et~al.}}]{Fu2024surpassing}%
  \BibitemOpen
  \bibfield  {author} {\bibinfo {author} {\bibfnamefont {R.}~\bibnamefont {Fu}}, \bibinfo {author} {\bibfnamefont {Z.}~\bibnamefont {Su}}, \bibinfo {author} {\bibfnamefont {H.-S.}\ \bibnamefont {Zhong}}, \bibinfo {author} {\bibfnamefont {X.}~\bibnamefont {Zhao}}, \bibinfo {author} {\bibfnamefont {J.}~\bibnamefont {Zhang}}, \bibinfo {author} {\bibfnamefont {F.}~\bibnamefont {Pan}}, \bibinfo {author} {\bibfnamefont {P.}~\bibnamefont {Zhang}}, \bibinfo {author} {\bibfnamefont {X.}~\bibnamefont {Zhao}}, \bibinfo {author} {\bibfnamefont {M.-C.}\ \bibnamefont {Chen}}, \bibinfo {author} {\bibfnamefont {C.-Y.}\ \bibnamefont {Lu}},  \emph {et~al.},\ }in\ \href@noop {} {\emph {\bibinfo {booktitle} {2024 SC24: International Conference for High Performance Computing, Networking, Storage and Analysis SC}}}\ (\bibinfo {organization} {IEEE Computer Society},\ \bibinfo {year} {2024})\ pp.\ \bibinfo {pages} {1241--1260}\BibitemShut {NoStop}%
\bibitem [{\citenamefont {Zhao}\ \emph {et~al.}(2024)\citenamefont {Zhao}, \citenamefont {Zhong}, \citenamefont {Pan}, \citenamefont {Chen}, \citenamefont {Fu}, \citenamefont {Su}, \citenamefont {Xie}, \citenamefont {Zhao}, \citenamefont {Zhang}, \citenamefont {Ouyang} \emph {et~al.}}]{zhao2024leapfrogging}%
  \BibitemOpen
  \bibfield  {author} {\bibinfo {author} {\bibfnamefont {X.-H.}\ \bibnamefont {Zhao}}, \bibinfo {author} {\bibfnamefont {H.-S.}\ \bibnamefont {Zhong}}, \bibinfo {author} {\bibfnamefont {F.}~\bibnamefont {Pan}}, \bibinfo {author} {\bibfnamefont {Z.-H.}\ \bibnamefont {Chen}}, \bibinfo {author} {\bibfnamefont {R.}~\bibnamefont {Fu}}, \bibinfo {author} {\bibfnamefont {Z.}~\bibnamefont {Su}}, \bibinfo {author} {\bibfnamefont {X.}~\bibnamefont {Xie}}, \bibinfo {author} {\bibfnamefont {C.}~\bibnamefont {Zhao}}, \bibinfo {author} {\bibfnamefont {P.}~\bibnamefont {Zhang}}, \bibinfo {author} {\bibfnamefont {W.}~\bibnamefont {Ouyang}},  \emph {et~al.},\ }\href@noop {} {\bibfield  {journal} {\bibinfo  {journal} {National Science Review}\ ,\ \bibinfo {pages} {nwae317}} (\bibinfo {year} {2024})}\BibitemShut {NoStop}%
\bibitem [{\citenamefont {Guan}\ \emph {et~al.}(2024)\citenamefont {Guan}, \citenamefont {Zhou}, \citenamefont {Albarr{\'a}n-Arriagada}, \citenamefont {Chen}, \citenamefont {Solano}, \citenamefont {Hegade},\ and\ \citenamefont {Huang}}]{guan2024single}%
  \BibitemOpen
  \bibfield  {author} {\bibinfo {author} {\bibfnamefont {H.}~\bibnamefont {Guan}}, \bibinfo {author} {\bibfnamefont {F.}~\bibnamefont {Zhou}}, \bibinfo {author} {\bibfnamefont {F.}~\bibnamefont {Albarr{\'a}n-Arriagada}}, \bibinfo {author} {\bibfnamefont {X.}~\bibnamefont {Chen}}, \bibinfo {author} {\bibfnamefont {E.}~\bibnamefont {Solano}}, \bibinfo {author} {\bibfnamefont {N.~N.}\ \bibnamefont {Hegade}}, \ and\ \bibinfo {author} {\bibfnamefont {H.-L.}\ \bibnamefont {Huang}},\ }\href@noop {} {\bibfield  {journal} {\bibinfo  {journal} {Quantum Science and Technology}\ }\textbf {\bibinfo {volume} {10}},\ \bibinfo {pages} {015006} (\bibinfo {year} {2024})}\BibitemShut {NoStop}%
\bibitem [{\citenamefont {Huang}\ \emph {et~al.}(2021)\citenamefont {Huang}, \citenamefont {Du}, \citenamefont {Gong}, \citenamefont {Zhao}, \citenamefont {Wu}, \citenamefont {Wang}, \citenamefont {Li}, \citenamefont {Liang}, \citenamefont {Lin}, \citenamefont {Xu} \emph {et~al.}}]{huang2021experimental}%
  \BibitemOpen
  \bibfield  {author} {\bibinfo {author} {\bibfnamefont {H.-L.}\ \bibnamefont {Huang}}, \bibinfo {author} {\bibfnamefont {Y.}~\bibnamefont {Du}}, \bibinfo {author} {\bibfnamefont {M.}~\bibnamefont {Gong}}, \bibinfo {author} {\bibfnamefont {Y.}~\bibnamefont {Zhao}}, \bibinfo {author} {\bibfnamefont {Y.}~\bibnamefont {Wu}}, \bibinfo {author} {\bibfnamefont {C.}~\bibnamefont {Wang}}, \bibinfo {author} {\bibfnamefont {S.}~\bibnamefont {Li}}, \bibinfo {author} {\bibfnamefont {F.}~\bibnamefont {Liang}}, \bibinfo {author} {\bibfnamefont {J.}~\bibnamefont {Lin}}, \bibinfo {author} {\bibfnamefont {Y.}~\bibnamefont {Xu}},  \emph {et~al.},\ }\href@noop {} {\bibfield  {journal} {\bibinfo  {journal} {Physical Review Applied}\ }\textbf {\bibinfo {volume} {16}},\ \bibinfo {pages} {024051} (\bibinfo {year} {2021})}\BibitemShut {NoStop}%
\bibitem [{\citenamefont {Liu}\ \emph {et~al.}(2021)\citenamefont {Liu}, \citenamefont {Lim}, \citenamefont {Wood}, \citenamefont {Huang}, \citenamefont {Guo},\ and\ \citenamefont {Huang}}]{liu2021hybrid}%
  \BibitemOpen
  \bibfield  {author} {\bibinfo {author} {\bibfnamefont {J.}~\bibnamefont {Liu}}, \bibinfo {author} {\bibfnamefont {K.~H.}\ \bibnamefont {Lim}}, \bibinfo {author} {\bibfnamefont {K.~L.}\ \bibnamefont {Wood}}, \bibinfo {author} {\bibfnamefont {W.}~\bibnamefont {Huang}}, \bibinfo {author} {\bibfnamefont {C.}~\bibnamefont {Guo}}, \ and\ \bibinfo {author} {\bibfnamefont {H.-L.}\ \bibnamefont {Huang}},\ }\href@noop {} {\bibfield  {journal} {\bibinfo  {journal} {Science China Physics, Mechanics \& Astronomy}\ }\textbf {\bibinfo {volume} {64}},\ \bibinfo {pages} {290311} (\bibinfo {year} {2021})}\BibitemShut {NoStop}%
\bibitem [{\citenamefont {Gong}\ \emph {et~al.}(2023)\citenamefont {Gong}, \citenamefont {Huang}, \citenamefont {Wang}, \citenamefont {Guo}, \citenamefont {Li}, \citenamefont {Wu}, \citenamefont {Zhu}, \citenamefont {Zhao}, \citenamefont {Guo}, \citenamefont {Qian} \emph {et~al.}}]{gong2023quantum}%
  \BibitemOpen
  \bibfield  {author} {\bibinfo {author} {\bibfnamefont {M.}~\bibnamefont {Gong}}, \bibinfo {author} {\bibfnamefont {H.-L.}\ \bibnamefont {Huang}}, \bibinfo {author} {\bibfnamefont {S.}~\bibnamefont {Wang}}, \bibinfo {author} {\bibfnamefont {C.}~\bibnamefont {Guo}}, \bibinfo {author} {\bibfnamefont {S.}~\bibnamefont {Li}}, \bibinfo {author} {\bibfnamefont {Y.}~\bibnamefont {Wu}}, \bibinfo {author} {\bibfnamefont {Q.}~\bibnamefont {Zhu}}, \bibinfo {author} {\bibfnamefont {Y.}~\bibnamefont {Zhao}}, \bibinfo {author} {\bibfnamefont {S.}~\bibnamefont {Guo}}, \bibinfo {author} {\bibfnamefont {H.}~\bibnamefont {Qian}},  \emph {et~al.},\ }\href@noop {} {\bibfield  {journal} {\bibinfo  {journal} {Science Bulletin}\ }\textbf {\bibinfo {volume} {68}},\ \bibinfo {pages} {906} (\bibinfo {year} {2023})}\BibitemShut {NoStop}%
\end{thebibliography}%


\begin{thebibliography}{11}%
\makeatletter
\providecommand \@ifxundefined [1]{%
 \@ifx{#1\undefined}
}%
\providecommand \@ifnum [1]{%
 \ifnum #1\expandafter \@firstoftwo
 \else \expandafter \@secondoftwo
 \fi
}%
\providecommand \@ifx [1]{%
 \ifx #1\expandafter \@firstoftwo
 \else \expandafter \@secondoftwo
 \fi
}%
\providecommand \natexlab [1]{#1}%
\providecommand \enquote  [1]{``#1''}%
\providecommand \bibnamefont  [1]{#1}%
\providecommand \bibfnamefont [1]{#1}%
\providecommand \citenamefont [1]{#1}%
\providecommand \href@noop [0]{\@secondoftwo}%
\providecommand \href [0]{\begingroup \@sanitize@url \@href}%
\providecommand \@href[1]{\@@startlink{#1}\@@href}%
\providecommand \@@href[1]{\endgroup#1\@@endlink}%
\providecommand \@sanitize@url [0]{\catcode `\\12\catcode `\$12\catcode `\&12\catcode `\#12\catcode `\^12\catcode `\_12\catcode `\%12\relax}%
\providecommand \@@startlink[1]{}%
\providecommand \@@endlink[0]{}%
\providecommand \url  [0]{\begingroup\@sanitize@url \@url }%
\providecommand \@url [1]{\endgroup\@href {#1}{\urlprefix }}%
\providecommand \urlprefix  [0]{URL }%
\providecommand \Eprint [0]{\href }%
\providecommand \doibase [0]{http://dx.doi.org/}%
\providecommand \selectlanguage [0]{\@gobble}%
\providecommand \bibinfo  [0]{\@secondoftwo}%
\providecommand \bibfield  [0]{\@secondoftwo}%
\providecommand \translation [1]{[#1]}%
\providecommand \BibitemOpen [0]{}%
\providecommand \bibitemStop [0]{}%
\providecommand \bibitemNoStop [0]{.\EOS\space}%
\providecommand \EOS [0]{\spacefactor3000\relax}%
\providecommand \BibitemShut  [1]{\csname bibitem#1\endcsname}%
\let\auto@bib@innerbib\@empty
\bibitem [{\citenamefont {Koch}\ \emph {et~al.}(2007)\citenamefont {Koch}, \citenamefont {Yu}, \citenamefont {Gambetta}, \citenamefont {Houck}, \citenamefont {Schuster}, \citenamefont {Majer}, \citenamefont {Blais}, \citenamefont {Devoret}, \citenamefont {Girvin},\ and\ \citenamefont {Schoelkopf}}]{koch2007charge}%
  \BibitemOpen
  \bibfield  {author} {\bibinfo {author} {\bibfnamefont {J.}~\bibnamefont {Koch}}, \bibinfo {author} {\bibfnamefont {T.~M.}\ \bibnamefont {Yu}}, \bibinfo {author} {\bibfnamefont {J.}~\bibnamefont {Gambetta}}, \bibinfo {author} {\bibfnamefont {A.~A.}\ \bibnamefont {Houck}}, \bibinfo {author} {\bibfnamefont {D.~I.}\ \bibnamefont {Schuster}}, \bibinfo {author} {\bibfnamefont {J.}~\bibnamefont {Majer}}, \bibinfo {author} {\bibfnamefont {A.}~\bibnamefont {Blais}}, \bibinfo {author} {\bibfnamefont {M.~H.}\ \bibnamefont {Devoret}}, \bibinfo {author} {\bibfnamefont {S.~M.}\ \bibnamefont {Girvin}}, \ and\ \bibinfo {author} {\bibfnamefont {R.~J.}\ \bibnamefont {Schoelkopf}},\ }\href@noop {} {\bibfield  {journal} {\bibinfo  {journal} {Physical Review A—Atomic, Molecular, and Optical Physics}\ }\textbf {\bibinfo {volume} {76}},\ \bibinfo {pages} {042319} (\bibinfo {year} {2007})}\BibitemShut {NoStop}%
\bibitem [{\citenamefont {Ye}\ \emph {et~al.}(2021)\citenamefont {Ye}, \citenamefont {Cao}, \citenamefont {Wu}, \citenamefont {Chen}, \citenamefont {Zhu}, \citenamefont {Li}, \citenamefont {Chen}, \citenamefont {Gong}, \citenamefont {Zha}, \citenamefont {Huang} \emph {et~al.}}]{ye2021realization}%
  \BibitemOpen
  \bibfield  {author} {\bibinfo {author} {\bibfnamefont {Y.}~\bibnamefont {Ye}}, \bibinfo {author} {\bibfnamefont {S.}~\bibnamefont {Cao}}, \bibinfo {author} {\bibfnamefont {Y.}~\bibnamefont {Wu}}, \bibinfo {author} {\bibfnamefont {X.}~\bibnamefont {Chen}}, \bibinfo {author} {\bibfnamefont {Q.}~\bibnamefont {Zhu}}, \bibinfo {author} {\bibfnamefont {S.}~\bibnamefont {Li}}, \bibinfo {author} {\bibfnamefont {F.}~\bibnamefont {Chen}}, \bibinfo {author} {\bibfnamefont {M.}~\bibnamefont {Gong}}, \bibinfo {author} {\bibfnamefont {C.}~\bibnamefont {Zha}}, \bibinfo {author} {\bibfnamefont {H.-L.}\ \bibnamefont {Huang}},  \emph {et~al.},\ }\href@noop {} {\bibfield  {journal} {\bibinfo  {journal} {Chinese Physics Letters}\ }\textbf {\bibinfo {volume} {38}},\ \bibinfo {pages} {100301} (\bibinfo {year} {2021})}\BibitemShut {NoStop}%
\bibitem [{\citenamefont {Wu}\ \emph {et~al.}(2021)\citenamefont {Wu}, \citenamefont {Bao}, \citenamefont {Cao}, \citenamefont {Chen}, \citenamefont {Chen}, \citenamefont {Chen}, \citenamefont {Chung}, \citenamefont {Deng}, \citenamefont {Du}, \citenamefont {Fan} \emph {et~al.}}]{wu2021strong}%
  \BibitemOpen
  \bibfield  {author} {\bibinfo {author} {\bibfnamefont {Y.}~\bibnamefont {Wu}}, \bibinfo {author} {\bibfnamefont {W.-S.}\ \bibnamefont {Bao}}, \bibinfo {author} {\bibfnamefont {S.}~\bibnamefont {Cao}}, \bibinfo {author} {\bibfnamefont {F.}~\bibnamefont {Chen}}, \bibinfo {author} {\bibfnamefont {M.-C.}\ \bibnamefont {Chen}}, \bibinfo {author} {\bibfnamefont {X.}~\bibnamefont {Chen}}, \bibinfo {author} {\bibfnamefont {T.-H.}\ \bibnamefont {Chung}}, \bibinfo {author} {\bibfnamefont {H.}~\bibnamefont {Deng}}, \bibinfo {author} {\bibfnamefont {Y.}~\bibnamefont {Du}}, \bibinfo {author} {\bibfnamefont {D.}~\bibnamefont {Fan}},  \emph {et~al.},\ }\href@noop {} {\bibfield  {journal} {\bibinfo  {journal} {Physical Review Letters}\ }\textbf {\bibinfo {volume} {127}},\ \bibinfo {pages} {180501} (\bibinfo {year} {2021})}\BibitemShut {NoStop}%
\bibitem [{\citenamefont {Yurke}\ \emph {et~al.}(1996)\citenamefont {Yurke}, \citenamefont {Roukes}, \citenamefont {Movshovich},\ and\ \citenamefont {Pargellis}}]{yurke1996low}%
  \BibitemOpen
  \bibfield  {author} {\bibinfo {author} {\bibfnamefont {B.}~\bibnamefont {Yurke}}, \bibinfo {author} {\bibfnamefont {M.}~\bibnamefont {Roukes}}, \bibinfo {author} {\bibfnamefont {R.}~\bibnamefont {Movshovich}}, \ and\ \bibinfo {author} {\bibfnamefont {A.}~\bibnamefont {Pargellis}},\ }\href@noop {} {\bibfield  {journal} {\bibinfo  {journal} {Applied physics letters}\ }\textbf {\bibinfo {volume} {69}},\ \bibinfo {pages} {3078} (\bibinfo {year} {1996})}\BibitemShut {NoStop}%
\bibitem [{\citenamefont {Ranadive}\ \emph {et~al.}(2022)\citenamefont {Ranadive}, \citenamefont {Esposito}, \citenamefont {Planat}, \citenamefont {Bonet}, \citenamefont {Naud}, \citenamefont {Buisson}, \citenamefont {Guichard},\ and\ \citenamefont {Roch}}]{ranadive2022kerr}%
  \BibitemOpen
  \bibfield  {author} {\bibinfo {author} {\bibfnamefont {A.}~\bibnamefont {Ranadive}}, \bibinfo {author} {\bibfnamefont {M.}~\bibnamefont {Esposito}}, \bibinfo {author} {\bibfnamefont {L.}~\bibnamefont {Planat}}, \bibinfo {author} {\bibfnamefont {E.}~\bibnamefont {Bonet}}, \bibinfo {author} {\bibfnamefont {C.}~\bibnamefont {Naud}}, \bibinfo {author} {\bibfnamefont {O.}~\bibnamefont {Buisson}}, \bibinfo {author} {\bibfnamefont {W.}~\bibnamefont {Guichard}}, \ and\ \bibinfo {author} {\bibfnamefont {N.}~\bibnamefont {Roch}},\ }\href@noop {} {\bibfield  {journal} {\bibinfo  {journal} {Nature communications}\ }\textbf {\bibinfo {volume} {13}},\ \bibinfo {pages} {1737} (\bibinfo {year} {2022})}\BibitemShut {NoStop}%
\bibitem [{\citenamefont {Arute}\ \emph {et~al.}(2019)\citenamefont {Arute}, \citenamefont {Arya}, \citenamefont {Babbush}, \citenamefont {Bacon}, \citenamefont {Bardin}, \citenamefont {Barends}, \citenamefont {Biswas}, \citenamefont {Boixo}, \citenamefont {Brandao}, \citenamefont {Buell} \emph {et~al.}}]{arute2019quantum}%
  \BibitemOpen
  \bibfield  {author} {\bibinfo {author} {\bibfnamefont {F.}~\bibnamefont {Arute}}, \bibinfo {author} {\bibfnamefont {K.}~\bibnamefont {Arya}}, \bibinfo {author} {\bibfnamefont {R.}~\bibnamefont {Babbush}}, \bibinfo {author} {\bibfnamefont {D.}~\bibnamefont {Bacon}}, \bibinfo {author} {\bibfnamefont {J.~C.}\ \bibnamefont {Bardin}}, \bibinfo {author} {\bibfnamefont {R.}~\bibnamefont {Barends}}, \bibinfo {author} {\bibfnamefont {R.}~\bibnamefont {Biswas}}, \bibinfo {author} {\bibfnamefont {S.}~\bibnamefont {Boixo}}, \bibinfo {author} {\bibfnamefont {F.~G.}\ \bibnamefont {Brandao}}, \bibinfo {author} {\bibfnamefont {D.~A.}\ \bibnamefont {Buell}},  \emph {et~al.},\ }\href@noop {} {\bibfield  {journal} {\bibinfo  {journal} {Nature}\ }\textbf {\bibinfo {volume} {574}},\ \bibinfo {pages} {505} (\bibinfo {year} {2019})}\BibitemShut {NoStop}%
\bibitem [{\citenamefont {Elder}\ \emph {et~al.}(2020)\citenamefont {Elder}, \citenamefont {Wang}, \citenamefont {Reinhold}, \citenamefont {Hann}, \citenamefont {Chou}, \citenamefont {Lester}, \citenamefont {Rosenblum}, \citenamefont {Frunzio}, \citenamefont {Jiang},\ and\ \citenamefont {Schoelkopf}}]{elder2020high}%
  \BibitemOpen
  \bibfield  {author} {\bibinfo {author} {\bibfnamefont {S.~S.}\ \bibnamefont {Elder}}, \bibinfo {author} {\bibfnamefont {C.~S.}\ \bibnamefont {Wang}}, \bibinfo {author} {\bibfnamefont {P.}~\bibnamefont {Reinhold}}, \bibinfo {author} {\bibfnamefont {C.~T.}\ \bibnamefont {Hann}}, \bibinfo {author} {\bibfnamefont {K.~S.}\ \bibnamefont {Chou}}, \bibinfo {author} {\bibfnamefont {B.~J.}\ \bibnamefont {Lester}}, \bibinfo {author} {\bibfnamefont {S.}~\bibnamefont {Rosenblum}}, \bibinfo {author} {\bibfnamefont {L.}~\bibnamefont {Frunzio}}, \bibinfo {author} {\bibfnamefont {L.}~\bibnamefont {Jiang}}, \ and\ \bibinfo {author} {\bibfnamefont {R.~J.}\ \bibnamefont {Schoelkopf}},\ }\href@noop {} {\bibfield  {journal} {\bibinfo  {journal} {Physical Review X}\ }\textbf {\bibinfo {volume} {10}},\ \bibinfo {pages} {011001} (\bibinfo {year} {2020})}\BibitemShut {NoStop}%
\bibitem [{\citenamefont {Campagne-Ibarcq}\ \emph {et~al.}(2013)\citenamefont {Campagne-Ibarcq}, \citenamefont {Flurin}, \citenamefont {Roch}, \citenamefont {Darson}, \citenamefont {Morfin}, \citenamefont {Mirrahimi}, \citenamefont {Devoret}, \citenamefont {Mallet},\ and\ \citenamefont {Huard}}]{campagne2013persistent}%
  \BibitemOpen
  \bibfield  {author} {\bibinfo {author} {\bibfnamefont {P.}~\bibnamefont {Campagne-Ibarcq}}, \bibinfo {author} {\bibfnamefont {E.}~\bibnamefont {Flurin}}, \bibinfo {author} {\bibfnamefont {N.}~\bibnamefont {Roch}}, \bibinfo {author} {\bibfnamefont {D.}~\bibnamefont {Darson}}, \bibinfo {author} {\bibfnamefont {P.}~\bibnamefont {Morfin}}, \bibinfo {author} {\bibfnamefont {M.}~\bibnamefont {Mirrahimi}}, \bibinfo {author} {\bibfnamefont {M.~H.}\ \bibnamefont {Devoret}}, \bibinfo {author} {\bibfnamefont {F.}~\bibnamefont {Mallet}}, \ and\ \bibinfo {author} {\bibfnamefont {B.}~\bibnamefont {Huard}},\ }\href@noop {} {\bibfield  {journal} {\bibinfo  {journal} {Physical Review X}\ }\textbf {\bibinfo {volume} {3}},\ \bibinfo {pages} {021008} (\bibinfo {year} {2013})}\BibitemShut {NoStop}%
\bibitem [{\citenamefont {Salath{\'e}}\ \emph {et~al.}(2018)\citenamefont {Salath{\'e}}, \citenamefont {Kurpiers}, \citenamefont {Karg}, \citenamefont {Lang}, \citenamefont {Andersen}, \citenamefont {Akin}, \citenamefont {Krinner}, \citenamefont {Eichler},\ and\ \citenamefont {Wallraff}}]{salathe2018low}%
  \BibitemOpen
  \bibfield  {author} {\bibinfo {author} {\bibfnamefont {Y.}~\bibnamefont {Salath{\'e}}}, \bibinfo {author} {\bibfnamefont {P.}~\bibnamefont {Kurpiers}}, \bibinfo {author} {\bibfnamefont {T.}~\bibnamefont {Karg}}, \bibinfo {author} {\bibfnamefont {C.}~\bibnamefont {Lang}}, \bibinfo {author} {\bibfnamefont {C.~K.}\ \bibnamefont {Andersen}}, \bibinfo {author} {\bibfnamefont {A.}~\bibnamefont {Akin}}, \bibinfo {author} {\bibfnamefont {S.}~\bibnamefont {Krinner}}, \bibinfo {author} {\bibfnamefont {C.}~\bibnamefont {Eichler}}, \ and\ \bibinfo {author} {\bibfnamefont {A.}~\bibnamefont {Wallraff}},\ }\href@noop {} {\bibfield  {journal} {\bibinfo  {journal} {Physical Review Applied}\ }\textbf {\bibinfo {volume} {9}},\ \bibinfo {pages} {034011} (\bibinfo {year} {2018})}\BibitemShut {NoStop}%
\bibitem [{\citenamefont {Yan}\ \emph {et~al.}(2019)\citenamefont {Yan}, \citenamefont {Zhang}, \citenamefont {Gong}, \citenamefont {Wu}, \citenamefont {Zheng}, \citenamefont {Li}, \citenamefont {Wang}, \citenamefont {Liang}, \citenamefont {Lin}, \citenamefont {Xu} \emph {et~al.}}]{yan2019strongly}%
  \BibitemOpen
  \bibfield  {author} {\bibinfo {author} {\bibfnamefont {Z.}~\bibnamefont {Yan}}, \bibinfo {author} {\bibfnamefont {Y.-R.}\ \bibnamefont {Zhang}}, \bibinfo {author} {\bibfnamefont {M.}~\bibnamefont {Gong}}, \bibinfo {author} {\bibfnamefont {Y.}~\bibnamefont {Wu}}, \bibinfo {author} {\bibfnamefont {Y.}~\bibnamefont {Zheng}}, \bibinfo {author} {\bibfnamefont {S.}~\bibnamefont {Li}}, \bibinfo {author} {\bibfnamefont {C.}~\bibnamefont {Wang}}, \bibinfo {author} {\bibfnamefont {F.}~\bibnamefont {Liang}}, \bibinfo {author} {\bibfnamefont {J.}~\bibnamefont {Lin}}, \bibinfo {author} {\bibfnamefont {Y.}~\bibnamefont {Xu}},  \emph {et~al.},\ }\href@noop {} {\bibfield  {journal} {\bibinfo  {journal} {Science}\ }\textbf {\bibinfo {volume} {364}},\ \bibinfo {pages} {753} (\bibinfo {year} {2019})}\BibitemShut {NoStop}%
\bibitem [{\citenamefont {Zhu}\ \emph {et~al.}(2022)\citenamefont {Zhu}, \citenamefont {Cao}, \citenamefont {Chen}, \citenamefont {Chen}, \citenamefont {Chen}, \citenamefont {Chung}, \citenamefont {Deng}, \citenamefont {Du}, \citenamefont {Fan}, \citenamefont {Gong} \emph {et~al.}}]{zhu2022quantum}%
  \BibitemOpen
  \bibfield  {author} {\bibinfo {author} {\bibfnamefont {Q.}~\bibnamefont {Zhu}}, \bibinfo {author} {\bibfnamefont {S.}~\bibnamefont {Cao}}, \bibinfo {author} {\bibfnamefont {F.}~\bibnamefont {Chen}}, \bibinfo {author} {\bibfnamefont {M.-C.}\ \bibnamefont {Chen}}, \bibinfo {author} {\bibfnamefont {X.}~\bibnamefont {Chen}}, \bibinfo {author} {\bibfnamefont {T.-H.}\ \bibnamefont {Chung}}, \bibinfo {author} {\bibfnamefont {H.}~\bibnamefont {Deng}}, \bibinfo {author} {\bibfnamefont {Y.}~\bibnamefont {Du}}, \bibinfo {author} {\bibfnamefont {D.}~\bibnamefont {Fan}}, \bibinfo {author} {\bibfnamefont {M.}~\bibnamefont {Gong}},  \emph {et~al.},\ }\href@noop {} {\bibfield  {journal} {\bibinfo  {journal} {Science Bulletin}\ }\textbf {\bibinfo {volume} {67}},\ \bibinfo {pages} {240} (\bibinfo {year} {2022})}\BibitemShut {NoStop}%
\end{thebibliography}%

\end{document}


\title{Supplemental Material for \\ ``Establishing a New Benchmark in Quantum Computational Advantage with 105-qubit \textit{Zuchongzhi}~3.0 Processor''}

\author{Dongxin Gao} 
\thanks{These authors contributed equally to this work.}
\HFRC
\SHRC
\HFNL
\author{Daojin Fan} 
\thanks{These authors contributed equally to this work.}
\HFRC
\SHRC
\HFNL
\author{Chen Zha}
\thanks{These authors contributed equally to this work.}
\HFRC
\SHRC
\HFNL
\author{Jiahao Bei}
\SHRC
\author{Guoqing Cai}
\SHRC
\author{Jianbin Cai}
\HFRC
\SHRC
\HFNL
\author{Sirui Cao}
\HFRC
\SHRC
\HFNL
\author{Xiangdong Zeng}
\HFRC
\author{Fusheng Chen}
\HFRC
\SHRC
\HFNL
\author{Jiang Chen}
\SHRC
% ...

\date{\today}

\maketitle

\setcounter{section}{0}
\renewcommand{\thefigure}{S\arabic{figure}}	
\renewcommand{\thetable}{S\arabic{table}}	
\renewcommand{\theequation}{S\arabic{equation}}	
\setcounter{figure}{0}
\setcounter{table}{0}
\setcounter{equation}{0}

\section{Device information}

\subsection{Quantum processor}
\textit{Zuchongzhi} 3.0 is a superconducting programmable quantum processor, comprising  of $105$ Transmon qubits~\cite{koch2007charge} and $182$ tunable couplers~\cite{ye2021realization}, arranged in 15 rows and 7 columns, as depicted in the main text. Each qubit exhibits an anhamonicity of approximately -230 MHz, with a control line for XY gates and iSWAP-like gates, except for the qubits $Q_{022}$ and $Q_{037}$, whose frequencies are non-tunable. For the precise measurement of qubit states, a readout resonator is individually coupled to each qubit, and shares a bandpass filter with the other six resonators to suppress the Purcell effect. 
Additionally, each coupler is equipped with a control line, allowing for the fine-tuning of the coupling strength between neighboring qubits.

The fabrication process closely follows that of \textit{Zuchongzhi} 2.0~\cite{wu2021strong},  with a key enhancement being the employment of tantalum film for the patterning of base wirings on the top substrate.

\subsection{Control electronics and Cryogenic wiring}
\begin{figure}[!htbp]
\begin{center}
\includegraphics[width=1.0\linewidth]{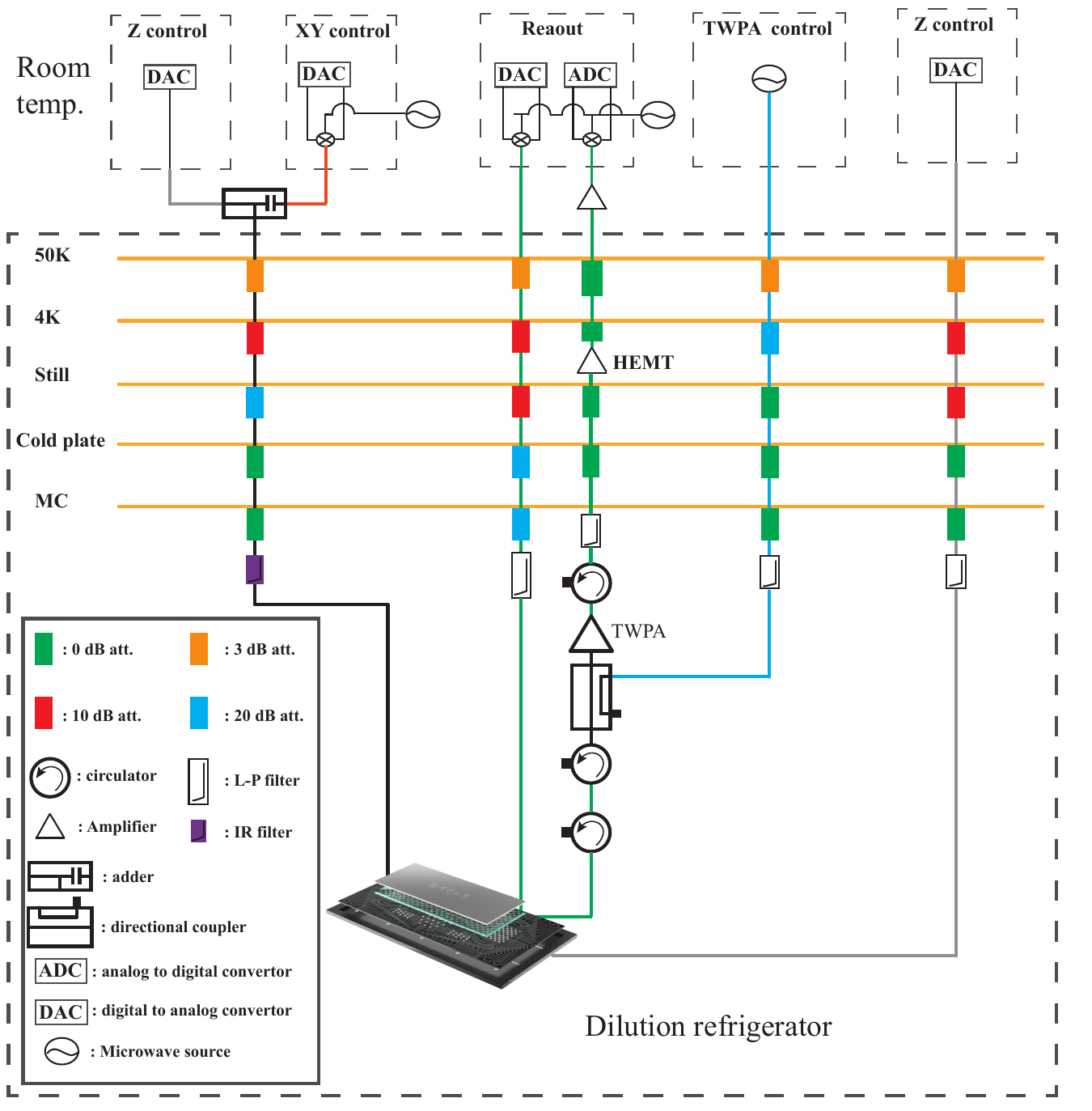}
\end{center}
\caption{\textbf{The schematic diagram of control electronics and wiring configuration}. Within the dilution refrigerator, attenuators and filters are installed at various plate to reduce the amplitude and noise of signals. For signal amplification, we employ Traveling-Wave Parametric Amplifiers (TWPAs), High Electron Mobility Transistors (HEMTs), and room-temperature amplifiers to enhance the readout signals. Directional couplers and circulators are utilized to prevent unwanted microwave coupling. At room temperature, the Z control pulse and XY control microwave are combined via an adder and the microwave sources are employed to pump the TWPA. }
\label{fig.wire_setup}
\end{figure}

The experimental wiring setup for qubit/coupler controls and frequency-multiplexed readouts is shown in Fig.~\ref{fig.wire_setup}. 

The cryogenic wiring used in this experiment is a modular high-density cable system. Above the mixture chamber plate, the dilution system consists of $2\times 12$ semi-rigid cables assembled in a modular fashion. In the mixture chamber plate, modular noise filters and thinner flexible phase-stable cables are used to ensure optimal signal transmission and noise reduction. Each layer is also equipped with modular attenuators, allowing for signal and noise damping. The modular design and the reduced diameter of cables allow for a large number of cables to be installed within the refrigerator. 

We combined the XY control lines and Z control lines at room temperature using combiners before connecting them to the cables inside the refrigerator, which further reduced the number of cables needed for the quantum chip. The refrigerator used in this experiment is equipped with a total of 504 cables, of which 332 cables were used in this experiment.

The control and readout setup employed in this experiment is the same as that of \textit{Zuchongzhi} 2.0. However, we have also made significant upgrades: the near-quantum-limited amplifier for readout has been replaced with a traveling-wave parametric amplifier (TWPA) \cite{yurke1996low,ranadive2022kerr}, and the configuration of the circulators has been correspondingly optimized. Additionally, a specific bandpass filter is added in front of the room-temperature amplifier to prevent the TWPA pump signal from infiltrating the ADC modules.

The room temperature electronic equipment in this experiment includes 407 DAC channels, 15 ADC modules, and 21 microwave source channels. The reduction in the number of microwave source channels is due to the use of microwave amplification modules, which allow a single microwave source channel to provide the local signals needed for 32 DAC channels.

\section{system calibration}
\begin{figure*}[!htbp]
\begin{center}
\includegraphics[width=1.0\linewidth]{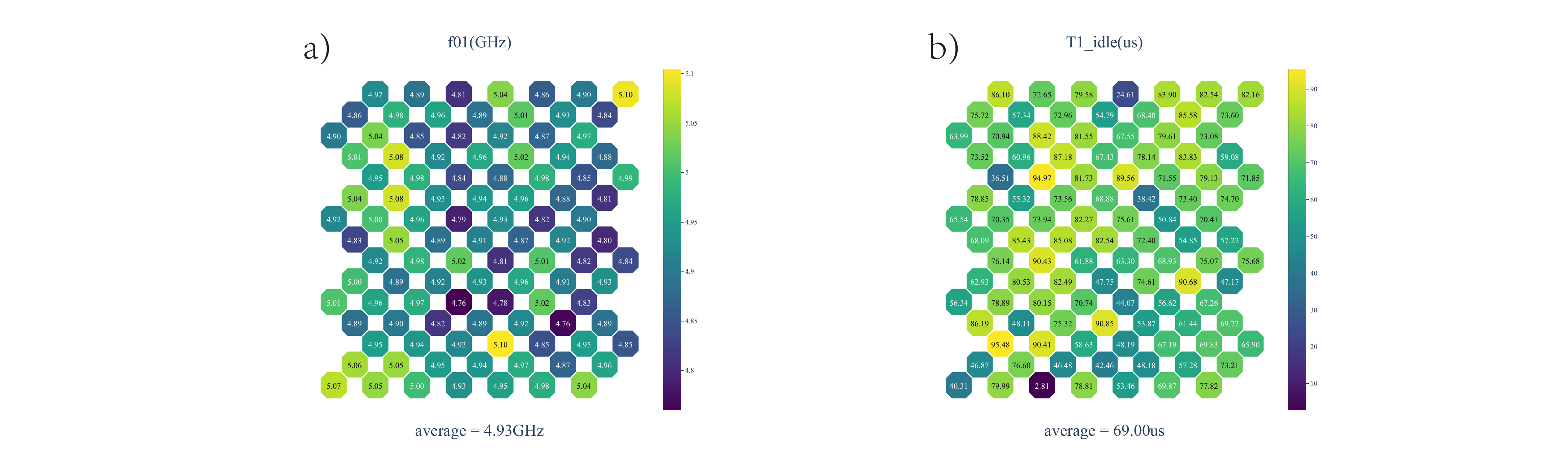}
\end{center}
\caption{\textbf{Typical parameter distribution of the 105-Qubit in the \textit{Zuchongzhi} 3.0}}
\label{fig.sys_parameters_105q}
\end{figure*}

The basic calibration encompasses a suite of experiments designed to ascertain the fundamental properties of the quantum processor, including readout frequencies, the responses of qubit frequencies to flux bias, XY crosstalk, and coherence times. The fundamental properties of the 105-qubit processor are list in the Fig.~\ref{fig.sys_parameters_105q}.

The calibration procedures for \textit{Zuchongzhi} 3.0 are analogous to those of \textit{Zuchongzhi} 2.0, as detailed in our previous work~\cite{wu2021strong}, and thus will not be reiterated in this manuscript. Instead, we will delve into the distinctive method applied in our random circuit sampling experiment.

\subsection{Idle Frequency Configuration for 83 qubits}

In our pursuit of quantum computational advantage with the processor's fundamental performance, we meticulously consider the fidelity of sampling, the complexity of the quantum circuits, and the inherent defects of both qubits and couplers. After a thorough evaluation, we have chosen a subset of 83 qubits for our random circuit sampling task, as shown in the main text. The other qubits are biased to their minimum frequencies to mitigate unwanted interactions with the active qubits. Additionally, we have turned off the coupling between the active qubits and those that are biased with couplers tuning.

For the avtive 83 qubits, we optimize the idle and interaction frequency arrangements, leveraging the basic calibration results of 83 qubits. During the idle frequency optimization, several critical factors are taken into account: coherence, two-level-system (TLS), residual coupling between qubits, XY-crosstalk, and Z pulse distortions. These considerations have enabled us to determine the optimal idle frequency configuration, as depicted in Fig~\ref{fig.sys_parameters_83} (b).

However, we observe that the $T_1$s performance at the idle frequencies sometimes exhibit significant variations throughout the experiment, attributed to the effects of the TLS. When qubits are subject to the influence of the TLS, we measure their $T_1$s near the idle frequencies and subsequently retune the idle frequencies of the affected qubits to mitigate the impact of the TLS.

With the stable and high-performance idle frequency configuration for 83 qubits, we proceed to calibrate the gate operations and readout processes employed in the random sampling experiment.

\subsection{Gate and Readout Calibration}

\subsubsection{Single-Qubit Gate Calibration}
After the highly precise calibration of the diving frequency, driving amplitude, and drag alpha for the single-qubit gate, we are enabled to realize a single-qubit gate with remarkably low control errors. However, the close adjacency of qubit control lines engenders crosstalk within the control signals. Consequently, this crosstalk induces discrepancies in single-qubit gate when they are carried out simultaneously as compared to when they are performed individually. Although the XY-crosstalk can be ameliorated by modulating the difference in idle frequencies between qubits, This concomitantly imposes more stringent constraints on the layout of idle frequencies, which becomes especially prominent as the qubit scale expands.

\begin{figure*}[!htbp]
\begin{center}
\includegraphics[width=1.0\linewidth]{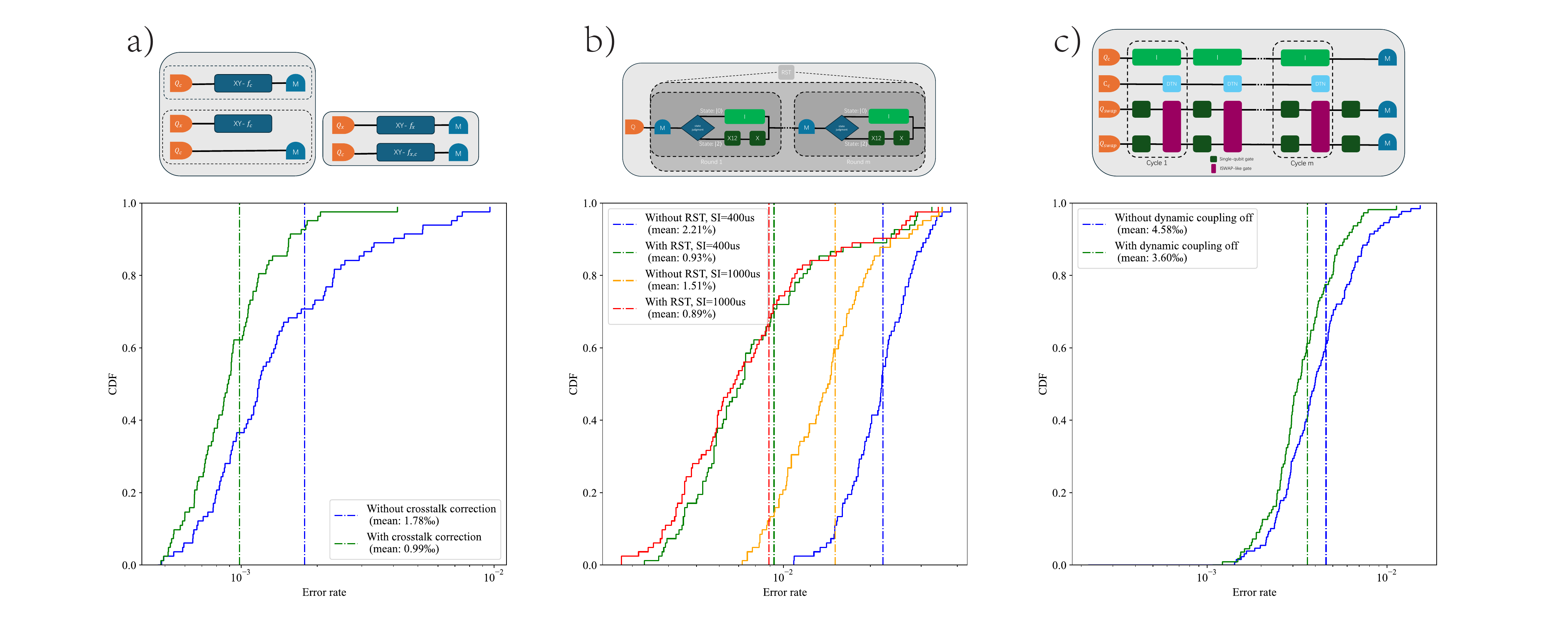}
\end{center}
\caption{\textbf{XY-crosstalk correction, active reset, dynamic coupling-off correction. a)} Top: The circuit diagram on the left is used for the rough correction of the amplitude correction coefficient  of the cancellation signal, while the circuit diagram on the right is used for the fine correction of the amplitude correction coefficient and phase of the cancellation signal. Bottom: CDF diagrams of single-qubit gate Pauli error with and without XY-crosstalk correction. \textbf{b)} Top: The schematic diagram of the active reset circuit. In our experiment, we carried out three rounds of active reset to ensure the qubit is reset back to the $\vert 0\rangle$ state. Bottom: CDF diagrams of readout error with and without active reset. \textbf{c)} Top: Circuit diagram of dynamic coupling-off. Bottom: CDF diagrams of two-qubit gate Pauli error with and without dynamic coupling-off. The vertical dotted lines in \textbf{a)}, \textbf{b)} and \textbf{c)} represent the average value. }
\label{fig.Sfig1}
\end{figure*}

To resolve this issue, we apply crosstalk cancellation signal to the compensated qubit $Q_c$, according to the driving waveform of the crosstalk qubit $Q_x$. 
The frequency of the cancellation waveform is the same as that of the driving waveform, while the amplitude is determined by the amplitude of the driving waveform, denoted as $A_c = A_x * \alpha_{cx}$, where $\alpha_{cx}$ represents the output ratio between the two channels. We obtained the initial value of $\alpha_{cx}$ by driving $Q_c$ with the frequency of $Q_c$ on the control lines of both the $Q_c$ and $Q_x$ respectively, and then measuring and comparing the periods of the corresponding Rabi oscillations utilizing the circuit depicted in the upper left corner of Fig. \ref{fig.Sfig1} (a). The additional phase $\delta_{\phi}$ of the cancellation waveform is extracted from the circuit shown in the upper right corner of Fig. \ref{fig.Sfig1} (a). We scan additional phase $\delta_{\phi}$ of the compensation microwave to observe the probability of $Q_c$ remaining in state $\vert 0\rangle$. The phase corresponding to the point with the highest probability is the optimal value. Subsequently, the precise value of $\alpha_{cx}$ is determined by Fine scan $\alpha_{cx}$. The bottom part of Fig. \ref{fig.Sfig1} (a) illustrates the comparison between the cases with and without XY-crosstalk correction, where the average Pauli error of Cross-Entropy Benchmarking (XEB)~\cite{arute2019quantum} is reduced from {$1.78\%$ to $0.99\%$}. 

\subsubsection{Readout Calibration}
We use a 0-2 readout method to improve the readout fidelity~\cite{elder2020high}, where an $X_{12}$ gate is applied to the qubit prior to measurement, preparing the qubit in either the $\ket{0}$ or $\ket{2}$ state. 
This method effectively reduces decoherence errors during readout but results in an increased sampling interval. 

To solve this issue, we implement an electronics-based active reset procedure~\cite{campagne2013persistent,salathe2018low}. Specifically, the qubit state is read at the beginning or end of the experiment, and a corresponding gate (denoted as CONDX) is subsequently applied based on the measured state. For the 0-2 readout method, when the qubit is measured in the $\ket{0}$ state, CONDX is the identity ($I$) gate, and when the qubit is measured in the $\ket{1}$ state, CONDX consists of the combination of $X_{12}+X_{01}$ gates. In Fig. \ref{fig.Sfig1} (b), the top graph represents the process of acitve reset, and the bottom graph is the CDF diagrams of readout error with and without active reset. Under the condition of sampling interval = $400\mu s$, the readout error is reduced from $2.21\%$ to $0.93\%$ after applying active reset.

To further improve the fidelity of the active reset process, we partition the phase space using a support vector machine (SVM) technique. However, this step introduces a small degree of initial state preparation errors, which will be calibrated in Section \ref{sec:state_praparation_error}.

 The readout error is obtained by simultaneously preparing all qubits in either $\ket{0}$ or $\ket{1}$, and measuring the corresponding error rates. To reduce correlated readout errors, we carefully tune the qubit frequencies during the readout process to avoid frequency collisions, achieving relatively low readout error rates. 

\subsubsection{Two-Qubit Gate Calibration}
The calibration and characterization process for the iSWAP-like gate adopted in this paper can be primarily summarized into the following steps:  
\begin{enumerate}
    \item We determine the swap frequency using an optimization algorithm, which primarily incorporates the $T_1$, $T_2$ of the qubits at the swap frequency, as well as frequency collisions with neighboring qubits. 
    \item With a fixed total gate time of 45ns, we can scan the two-dimensional map of the coupling strength $g$ and the detuning frequency $DTN$ required for aligning the frequencies of the two qubits. Here, we fix the detuning of one qubit and scan the detuning of the other to obtain initial values for $g$ and $DTN$.  
    \item The pulse distortion for the qubits is calibrated based on the initial detuning frequency $DTN$~\cite{yan2019strongly}. 
    \item We refine the optimal values of the coupling strength $g$ and the detuning parameter $DTN$. To enhance the sensitivity of the parameters, we meticulously adjust $g$ and $DTN$ by employing a sequence of odd-numbered iSWAP-like gates. 
    \item Due to periodic variations of two level system (TLS) and frequency collisions, the swap frequency optimized by the optimizer requires fine-tuning. We rescan the swap frequency, targeting the SPB value, to avoid frequency conflicts and TLS effects.  
    \item We perform a scan to determine the dynamic coupling-off point of neighboring qubits. During the implementation of the iSWAP-like gate, the relevant two qubits are detuned, which can lead to their unintended recoupling with other neighboring qubits. The recoupling can cause significant state leakage if their idle or operational frequencies are in close proximity. Therefore, we apply a $DTN$ to the neighboring coupler and scan for the optimal coupling-off point during gate operation. The specific circuit diagrams and the comparison of the two-qubit gate Pauli error with and without dynamic coupling-off are shown in Fig. \ref{fig.Sfig1} (c).
    \item We optimize the rising edge time of qubits and couplers. At this step, we scan the rising edge time of the waveforms of two qubits and the coupler. Similarly, we take the value of SPB as the benchmark to determine the optimal rising edge time.
    \item We employ XEB method to benchmark the parallel iSWAP-like gates performance.
\end{enumerate}

\begin{table*}[htb!]
    \centering
    \begin{tabular}{p{6cm}p{2.5cm}<{\centering}p{2.5cm}<{\centering}p{2.5cm}<{\centering}p{2.5cm}<{\centering}}
    \toprule
    Parameters& Mean& Median& Stdev.\\
    \toprule

    Qubit maximum frequency (GHz)& 5.039& 5.028& 0.071\\
    Qubit idle frequency (GHz)& 4.914& 4.920& 0.070\\ 
    Qubit anharmonicity  (MHz)& -229.4& -228.7& 3.8\\ 
    Readout drive frequency (GHz)& 6.408& 6.391& 0.100\\  
    $T_1$ at idle frequency ($\mu$s)& 71.6& 73.4& 13.8\\ 
    $T_2^{\text{CPMG}}$ at idle frequency ($\mu$s)& 57.8& 56.5& 15.3\\ 
    \hline
    Readout $e_{|0\rangle}$ simultaneous (\permil) & 4.97& 3.70 & 5.32\\  
    Readout $e_{|1\rangle}$ simultaneous (\permil) & 12.37 & 8.90& 8.99\\  
    1Q XEB $e_{1}$ simultaneous (\permil) & 0.97& 0.88& 0.41\\  
    1Q XEB $e_{1}$ purity simultaneous (\permil) & 0.63 & 0.54 & 0.27\\  
    2Q XEB $e_{2}$ simultaneous (\permil) & 3.75& 3.27& 1.60\\  
    2Q XEB $e_{2}$ purity simultaneous (\permil) & 3.59& 3.35& 1.50\\  

    \toprule
    \end{tabular}
    \caption{\textbf{Summary of system parameters of the 83-qubit subset.}}
    \label{tableSummary}
    \end{table*}

\begin{figure*}[!htbp]
\begin{center}
\includegraphics[width=1.0\linewidth]{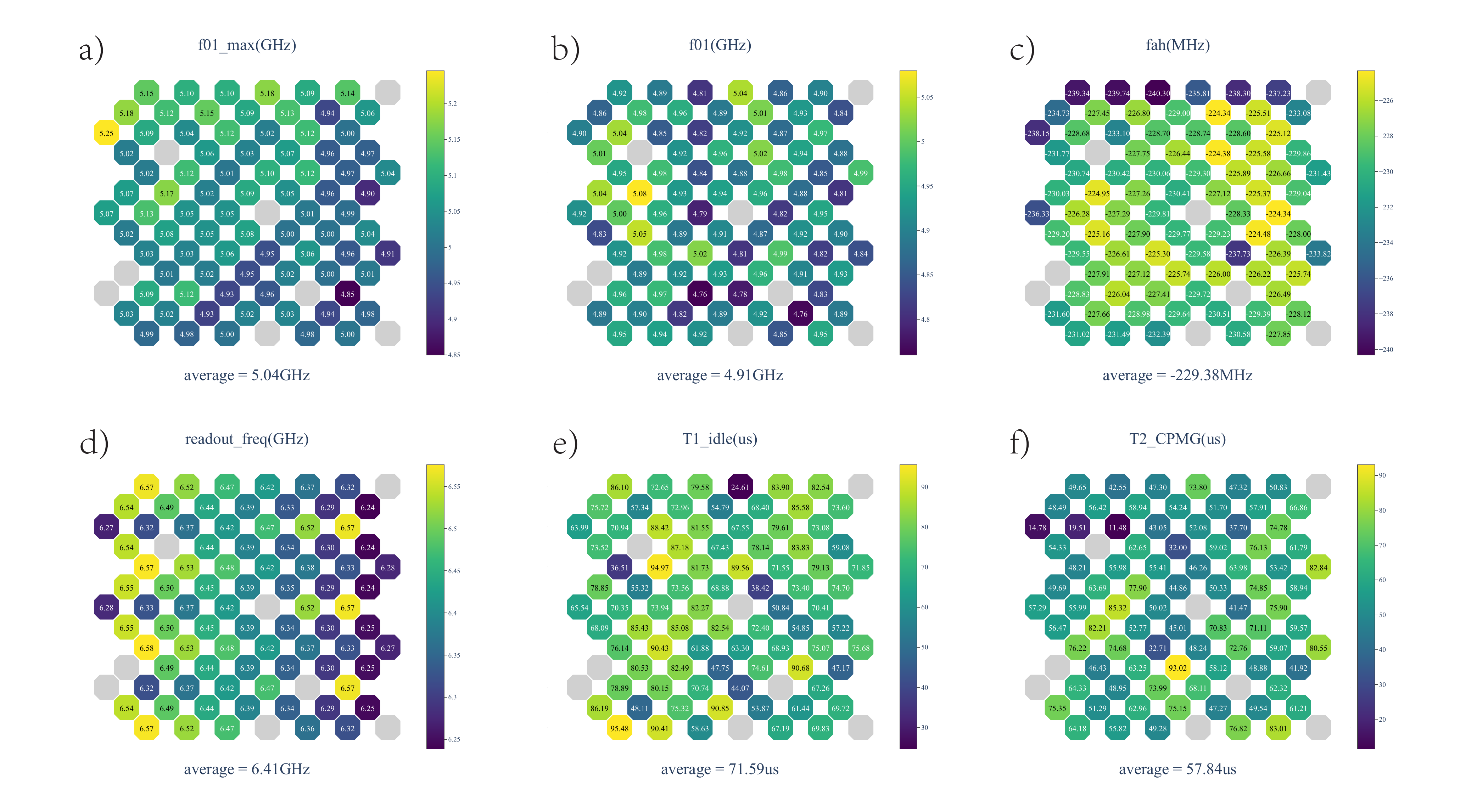}
\end{center}
\caption{\textbf{Typical parameter distribution of the 83-qubit subset in the \textit{Zuchongzhi} 3.0}}
\label{fig.sys_parameters_83}
\end{figure*}

The calibration results for the selected 83 qubits are presented in the Fig.~\ref{fig.sys_parameters_83}, and the detailed statistical data are presented in \ref{tableSummary}.

\subsection{Fine Calibration for Random Circuit Sampling Experiment}

%
Following the calibration of gate operations and readout processes, we achieve a high XEB fidelity for these operations. However, the quantum circuits used in the random circuit sampling experiment, which are executed in the sequence of ABCDCDAB, differ from the circuits employed in the XEB method, which are executed in the sequence of A. It leads to a deviation between experimental fidelity and predicted fidelity

Certain factors, including idle gate fidelities, pulse distortion of couplers, and state preparation errors, are not captured in the fidelities of XEB circuits but are reflected in the fidelities of the random circuits. Consequently, it is imperative that we calibrate for these factors to ensure the accuracy of random circuit sampling experiment.

It should be noticed that to closely align the calibrated effects of Z pulse crosstalk, dynamic coupling-off, and residual coupling with neighboring qubits, to those in the random circuit sampling experiment, we apply all waveforms to the qubits and couplers in the pattern during the calibration process.
 
\subsubsection{4-patch calibration}
The modular high-density cable configuration introduces significant Z pulse crosstalk between two Z control lines, resulting in substantial pulse distortion due to crosstalk.
The effect of Z pulse crosstalk can be mitigated by aligning calibration conditions as closely as possible with experimental conditions. However, the distortion associated with pulse crosstalk cannot be effectively reduced through standard calibration methods.

Due to the residual pulse distortion, the sequence of the pattern substantially alters the actual pulses applied to qubits and couplers during iSWAP-like gate operations. It leads to a discrepancy between the unitary matrix of the calibrated iSWAP-like gates and the one observed in the sampling experiment.

\begin{figure*}[!htbp]
\begin{center}
\includegraphics[width=1.0\linewidth]{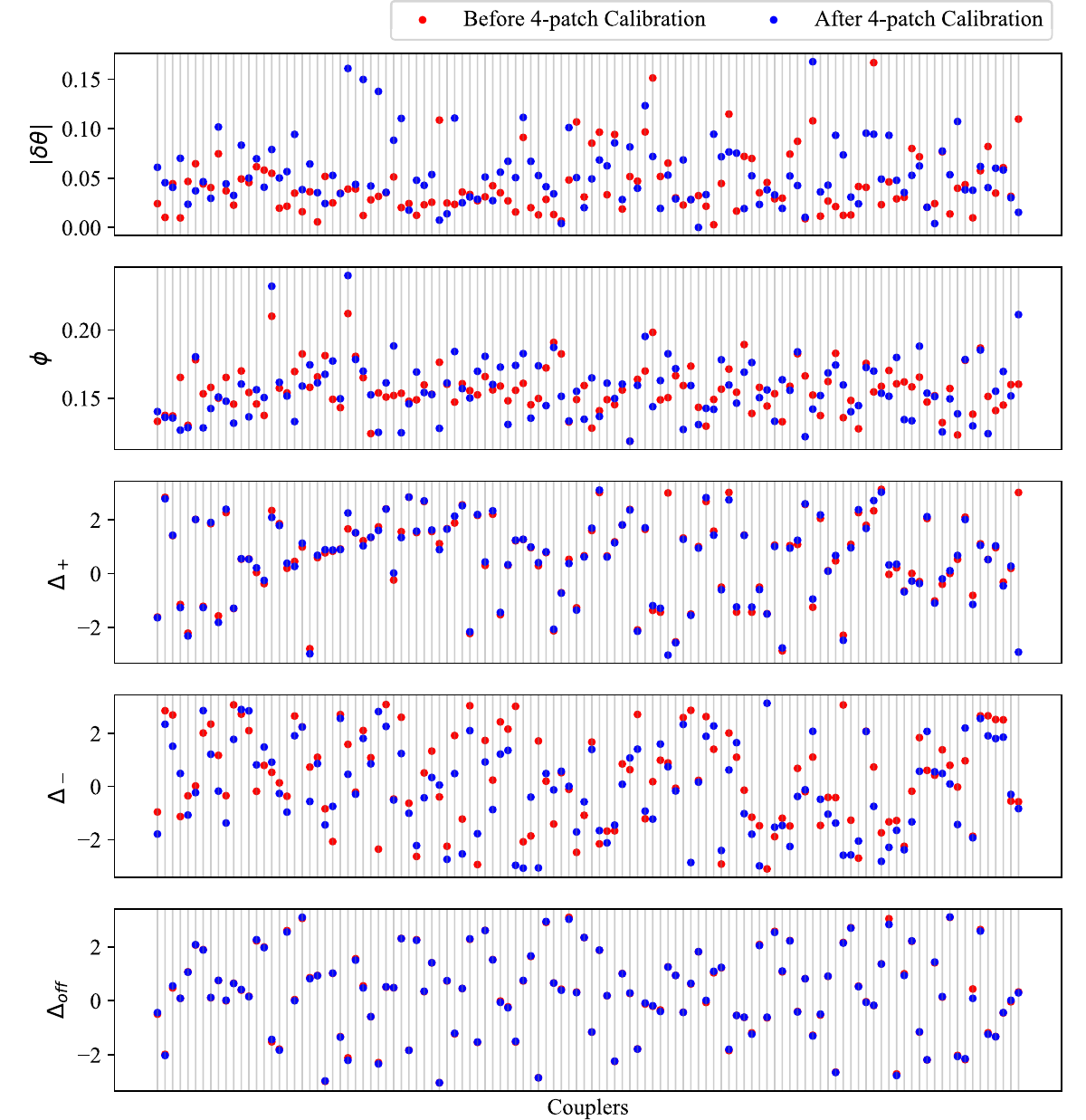}
\end{center}
\caption{\textbf{Comparison of the iSWAP-like gate parameters before and after 4-patch calibration in the Context of 83-qubit experiment.} The definitions of the five parameters in the figure can be found in Ref. \cite{zhu2022quantum}}
\label{fig.compare_U}
\end{figure*}

We employ the 4-patch calibration method, as detailed in the \textit{Zuchongzhi} 2.1~\cite{zhu2022quantum}, to optimize the parameters of the iSWAP-like gate used in our sampling experiment. This optimization is based on the original gate parameters derived from two-qubit XEB calibration. Figure~\ref{fig.compare_U} illustrates a comparison of the parameters before and after the 4-patch calibration. The 4-patch calibration brings the calibrated parameters closer to their actual values and aligns the experimental fidelities more closely with the predicted fidelities. Consequently, in all subsequent random circuit sampling experiments, unless otherwise specified, we perform the 4-patch calibration.

\subsubsection{Idle Gate Benchmarking and Calibration}
In the random circuit sampling experiment, we divided all the iSWAP-like gates into 4 groups of patterns and executed them alternately in a specific order. For instance, the order adopted in our paper is ABCDCDAB. As not all 83 qubits are engaged in every iSWAP-like gate pattern, certain qubits are designated to execute idle gates in various pattern. The idle gate time is 45 ns, which is the same as the total time of iSWAP-like gates. During the idle gates, the qubits are affected not only by their own decoherence but also by Z-crosstalk, dynamic coupling-off, and residual coupling with neighboring qubits. Consequently, we need to calibrate the errors of the idle gates in each pattern. 

\begin{figure*}[!htbp]
\begin{center}
\includegraphics[width=1.0\linewidth]{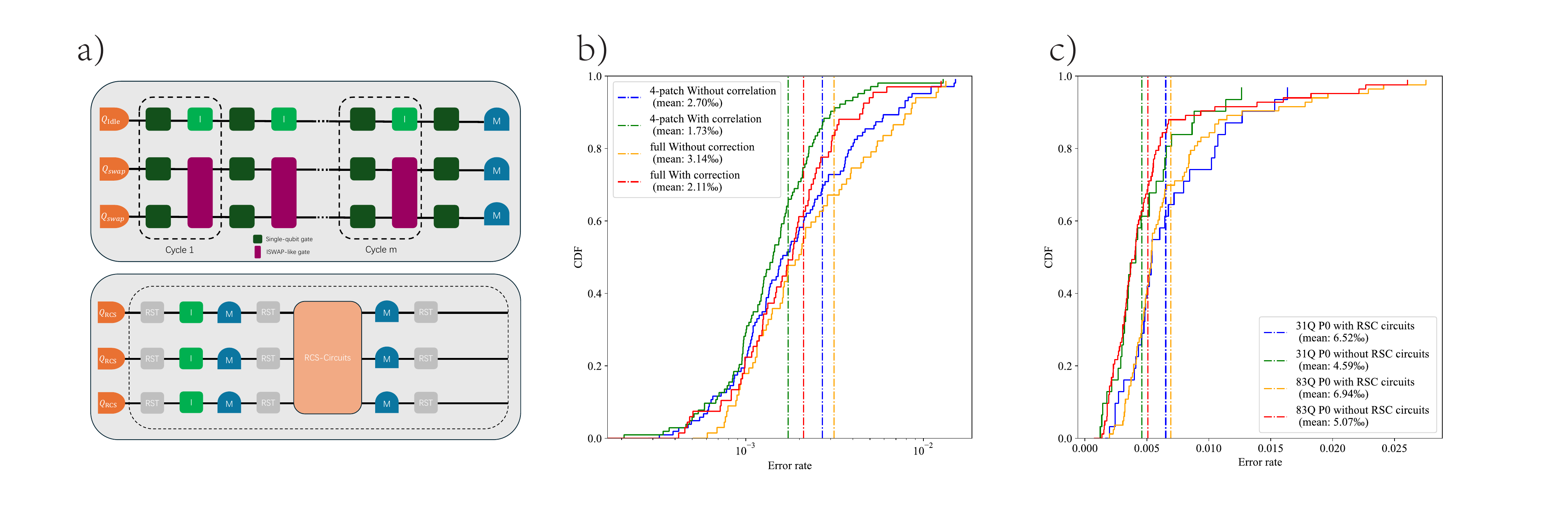}
\end{center}
\caption{\textbf{Extra error benchmarking in the RCS experiment. a)} Circuit diagrams for the benchmarking of the idle gate (top) and the readout error of P0 with RCS circuits (bottom). The single-qubit gates are chosen randomly from the set of {$\sqrt{X}, \sqrt{Y} , \sqrt{W}$}. \textbf{b)} CDF diagrams of the idle gate error with and without Z gate correction, in the 4-patch circuit and the full circuit respectively. \textbf{c)} CDF diagrams of P0 error with and without RCS-circuits, in 31-qubit and 83-qubit respectively. The vertical dotted lines in \textbf{b)} and \textbf{c)} represent the average value.}
\label{fig.Sfig2}
\end{figure*}

Firstly, we benchmark the fidelity of the idle gates utilizing the quantum circuit as shown in Fig. \ref{fig.Sfig2} (a), adopting a method akin to that used in single-qubit XEB experiments. Fig. \ref{fig.Sfig2} (b) illustrates that the fidelity of the idle gate significantly deviates from the fidelity estimated by SPB, due to the Z crosstalk, dynamic coupling-off, and residual coupling with neighboring qubits. 

To address the influence of the coupling with neighboring qubits, we re-scan for dynamic coupling-off and interaction frequencies. The qubit frequency shift can be corrected by applying Z gates as compensation. 
Balancing time and precision, we target the value of XBE under 100 cycles and scan the amplitude of Z gate to determine the optimal compensation amplitude. 

\subsubsection{Calibration of Coupler Distortion}
In our two-qubit gate calibration experiments, we achieve a notably high gate fidelity, even in the absence of corrections for pulse distortion of the couplers. However, during the random circuit sampling experiments, the experimental fidelity of each cycle is slightly lower than the fidelity predicted based on the number of gates. We hypothesize that this discrepancy arises because, in the two-qubit gate calibration, the relatively deep circuit leads to a stabilization of the coupler's pulse distortion. This stabilization allows the two-qubit gates at various cycles to be effectively represented by the same unitary matrix. Conversely, in the random circuit sampling experiments, the comparatively shallow circuit and the alternating sequence of four patterns result in two-qubit gates at different cycles being represented by distinct unitary matrices.

According to our hypothesis, it is imperative to correct the pulse distortion of the couplers. Given that we cannot directly measure the frequency variation of the coupler, we rely on indirect detection through associated qubits. Analogous to the pulse distortion correction for qubits, we can obtain the frequency response of the detection qubits at various times after the Z pulse of the coupler ends. By leveraging the relationship between the detuning amplitude of the coupler and the frequency shift of the detection qubits, we can deduce the distortion of the actual coupler voltage inversely. Once the response functions at different times are determined, we apply the correction method outlined in Ref.~\cite{yan2019strongly} to rectify the distortion. With the corrected Z pulses for the couplers, the experimental fidelity in each cycle of the random circuit sampling experiments aligns closely with the predicted fidelity.

\subsubsection{Calibration of State Preparation Errors}\label{sec:state_praparation_error}
Following the calibration of the idle gates and correction of coupler distortions, a consistent deviation between experimental and predicted fidelities remains, independent of circuit cycle. We attribute this discrepancy to state preparation errors. 

Since the effect of our current 02 readout reset scheme is based on the state of the qubit prior to the reset, the state preparation errors in the random circuit sampling process are different from those in the standard readout fidelity calibration.
To achieve a more realistic estimate of the state preparation errors, we adopted the circuit shown in Fig. \ref{fig.Sfig2} (a) to calibrate the state preparation error correction factor. In this circuit, except that the data of the first single shot is unavailable, we assume that the initial states in the subsequent experiments are consistent with those in the random circuit sampling experiment.

Due to the influence of the fact that the final states in the random circuit are not all 0, we can observe that the state $\ket{00\dots0}$ preparation fidelity obtained with the final state in the random circuit experiment is slightly lower than that of obtained with the the final state in the standard readout fidelity benchmarking as shown in Fig. \ref{fig.Sfig2} (c). Therefore, we need to consider this correction factor when calculating the estimated fidelity. 

As shown in Fig.~3 in the main text, after considering all the correction method mentioned above, the experiment and the prediction can match well. 

\section{RANDOM CIRCUIT SAMPLING of 31-Qubit 2-Patch Circuit}
\begin{figure*}[!htbp]
\begin{center}
\includegraphics[width=1.0\linewidth]{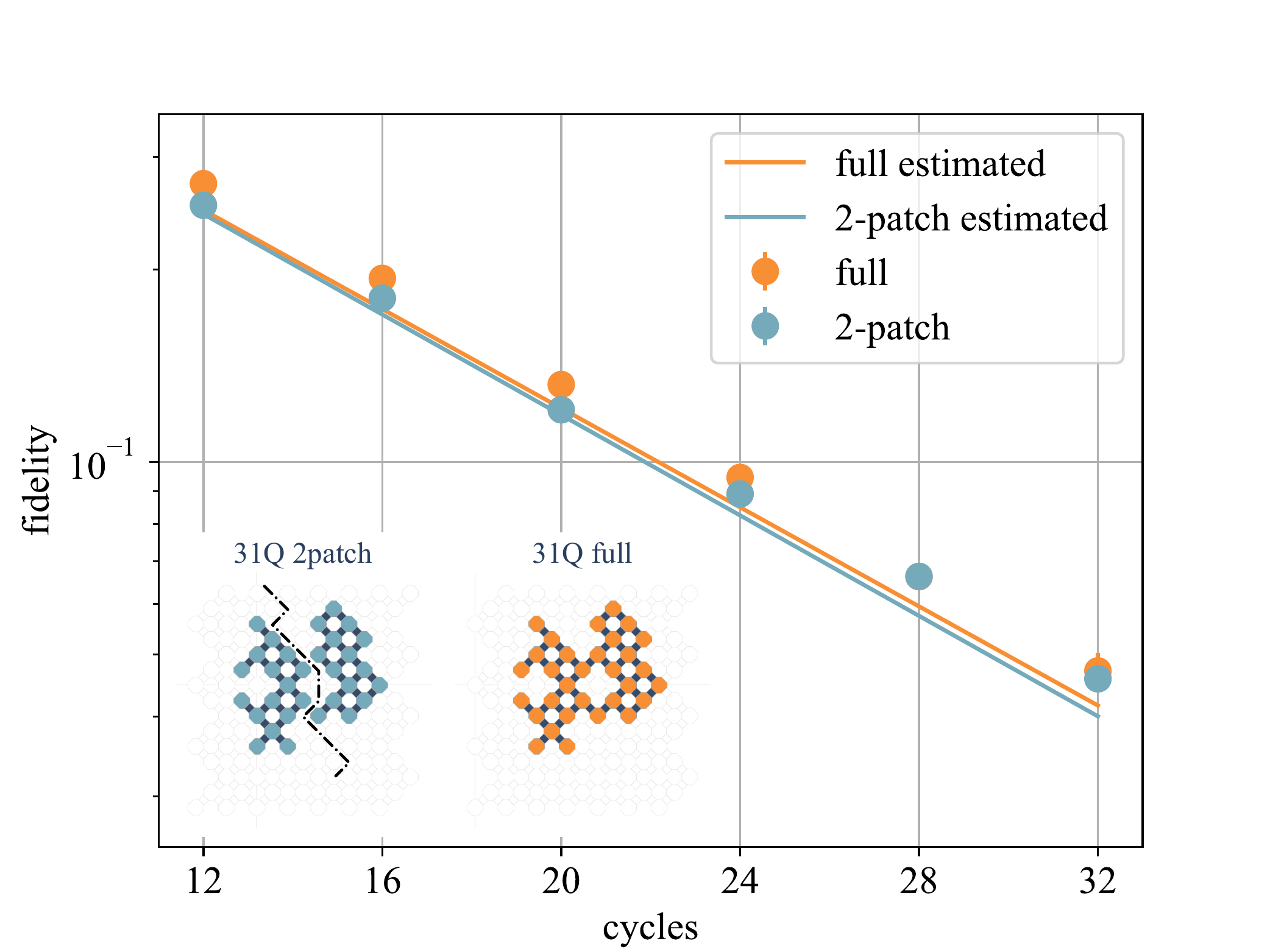}
\end{center}
\caption{\textbf{Experiment and estimate fidelity of random circuit sampling experiment for 31 qubits.} The green and blue dots respectively represent the experimental values of the 4-patch circuits and full circuits with 31 qubits over 12-36 cycles. The corresponding solid lines denote the estimated values for these circuits. The inserted topological diagram illustrates the specific configuration of 31 qubits.}
\label{fig.31qubit2patch}
\end{figure*}

We have also implemented the 2-patch version of the circuits, scaling from 12 to 32 cycles with 31 qubits each, and computed the linear XEB fidelities for the respective output bitstrings. The estimated and experimental fidelities for both the 2-patch circuits and the full circuits are depicted in Fig~\ref{fig.31qubit2patch}. The high degree of correspondence between the estimated and experimental results indicates that the discrete error model provides a reliable estimate of fidelity for both 2-patch circuits and full circuits when employing the 4-patch calibration method.

\section{random circuit sampling of 83-Qubit full Circuit}

\begin{figure*}[!htbp]
\begin{center}
\includegraphics[width=1.0\linewidth]{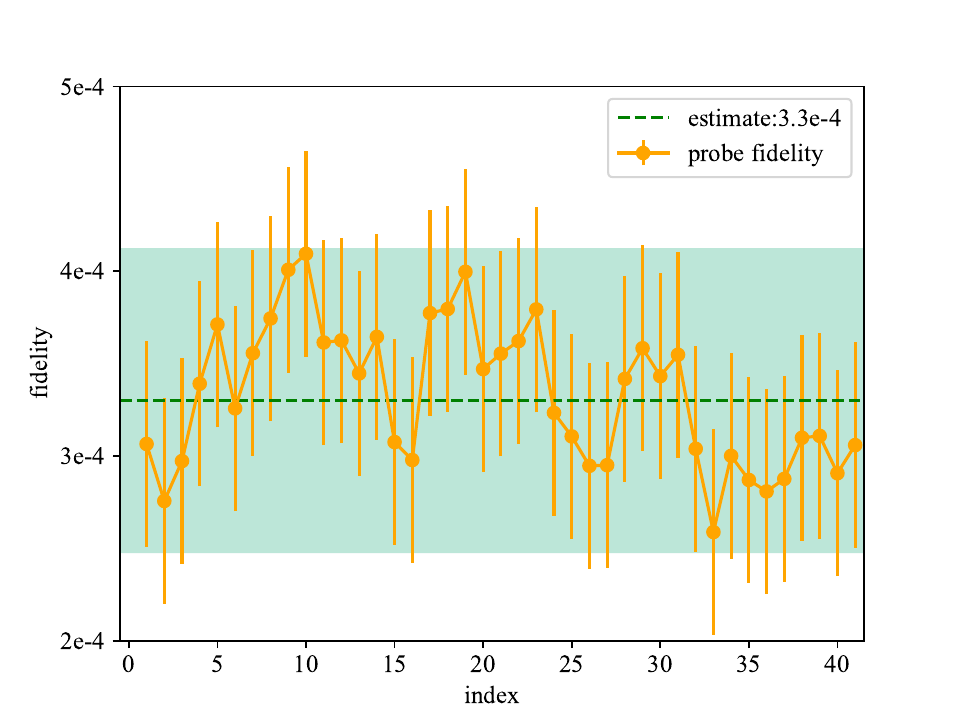}
\end{center}
\caption{\textbf{The fluctuation in probe fidelity over the course of the sampling process within the 83-qubit 32-cycle full circuit.} The dark green dotted line represents the estimated fidelity of 83-qubit 4-patch 32-cycle. The light green area indicates the range of plus or minus $25\%$ of the estimated fidelity. The yellow dots represent the experimental fidelity of 83-qubit, 4-patch and 32-cycle with half a million sampling bitstrings throughout the entire sampling process, and each dot is approximately spaced 2.4 hours apart.  It can be clearly seen that the experimental fidelity is within the light green area, demonstrating that our system is sufficiently stable during the sampling process. }
\label{fig.fidelity_4patch_probe}
\end{figure*}

In the random circuit sampling experiment conducted on the 83-qubit, 32-cycle full circuit, we sample a total of 410 million bitstrings, a process that spanned 91 hours. To ascertain the reliability of our experimental setup, we insert random circuit sampling experiments before and after every 10 million bitstrings within the 83-qubit full circuit, sampling half a million bitstrings on the 83-qubit, 4-patch, 32-cycle circuit. Subsequently, we monitor the fidelity variation over time, as illustrated in Fig.~\ref{fig.fidelity_4patch_probe}. From the results of the 83-qubit 4-patch 32-cycle circuit, it can be seen that the XEB fidelity was within the range of plus or minus $25\%$ of the estimated fidelity throughout the sampling period, which proves that our system had a sufficiently high fidelity and also verifies the validity of the data of the 83-qubit 32-cycle full circuit.

The computational complexity of random circuit sampling within an 83-qubit, 32-cycle system under various memory constraints is detailed in Table~\ref{Tab.complexity}.

\begin{table*}[!htbp]
\centering
\caption{Complexity under different memory constraints}
\renewcommand\arraystretch{1.2}
\begin{tabular}{|c|c|ccc|ccc|ccc|}
\hline
\multirow{2}{*}{\textbf{Experiment}}
& \multirow{2}{*}{\textbf{Fidelity}}
& \multicolumn{2}{c|}{\textbf{Memory Constraint: 9.2PB}}
& \multicolumn{2}{c|}{\textbf{Memory Constraint: 46.2PB}}
& \multicolumn{2}{c|}{\textbf{Memory Constraint: 762.2PB}} \\ \cline{3-8}
&
& \multicolumn{1}{c|}{\makecell{1 Amplitude}}
& \multicolumn{1}{c|}{\makecell{1 Million Noisy \\ Samples}}
& \multicolumn{1}{c|}{\makecell{1 Amplitude}}        
& \multicolumn{1}{c|}{\makecell{1 Million Noisy \\ Samples}}
& \multicolumn{1}{c|}{\makecell{1 Amplitude}}        
& \multicolumn{1}{c|}{\makecell{1 Million Noisy \\ Samples}} \\ \hline \hline
Sycamore-53-20   & $2.2 \times 10^{-3}$ & \multicolumn{1}{c|}{$7.2 \times 10^{18}$} & \multicolumn{1}{c|}{$6.5 \times 10^{16}$} & $9.5 \times 10^{18}$      & \multicolumn{1}{c|}{$5.9 \times 10^{16}$} & \multicolumn{1}{c|}{$5.9 \times 10^{18}$} & $6.1 \times 10^{16}$           \\ \hline
Zuchongzhi-56-20 & $6.6 \times 10^{-4}$ & \multicolumn{1}{c|}{$9.3 \times 10^{19}$} & \multicolumn{1}{c|}{$2.2 \times 10^{17}$} & $9.0 \times 10^{19}$    & \multicolumn{1}{c|}{$1.8 \times 10^{17}$} & \multicolumn{1}{c|}{$1.0 \times 10^{20}$} & $1.5 \times 10^{17}$              \\ \hline
Zuchongzhi-60-24 & $3.7 \times 10^{-4}$  & \multicolumn{1}{c|}{$3.2 \times 10^{21}$} & \multicolumn{1}{c|}{$1.6 \times 10^{19}$} & $3.0 \times 10^{21}$      & \multicolumn{1}{c|}{$1.0 \times 10^{19}$} & \multicolumn{1}{c|}{$3.0 \times 10^{21}$} & $2.3 \times 10^{18}$             \\ \hline
Sycamore-70-24    & $1.7 \times 10^{-3}$ & \multicolumn{1}{c|}{$1.7 \times 10^{25}$} & \multicolumn{1}{c|}{$8.2 \times 10^{25}$} & $4.1\times 10^{24}$      & \multicolumn{1}{c|}{$1.3 \times 10^{25}$} & \multicolumn{1}{c|}{$3.2 \times 10^{24}$} & $1.4 \times 10^{24}$           \\ \hline
Sycamore-67-32   & $1.5 \times 10^{-3}$ & \multicolumn{1}{c|}{$8.2 \times 10^{28}$} & \multicolumn{1}{c|}{$4.7 \times 10^{27}$} & $3.9\times 10^{27}$      & \multicolumn{1}{c|}{$2.6 \times 10^{26}$} & \multicolumn{1}{c|}{$1.3 \times 10^{26}$} & $9.6 \times 10^{24}$              \\ \hline
Zuchongzhi-83-32 & $2.5 \times 10^{-4}$ & \multicolumn{1}{c|}{$5.1 \times 10^{31}$} & \multicolumn{1}{c|}{$8.4 \times 10^{33}$} & $3.0\times 10^{30}$      & \multicolumn{1}{c|}{$1.3 \times 10^{32}$} & \multicolumn{1}{c|}{$1.3 \times 10^{29}$} & $7.5 \times 10^{31}$            \\ \hline
\end{tabular}
\label{Tab.complexity}
\end{table*}

\bibliography{references}